\documentclass[iop]{emulateapj}
\slugcomment{{\sc Accepted to ApJ:} July 9, 2016}
\usepackage{amssymb,amsmath}
\usepackage{graphicx}
\usepackage{natbib}
\usepackage{hyperref}

\begin{document}

\title{MC$^2$: Dynamical Analysis of the Merging Galaxy Cluster MACS J1149.5+2223}

\shorttitle{MACS J1149 Dynamical Analysis}

\author{Nathan Golovich\altaffilmark{1},
William A. Dawson\altaffilmark{2},
David Wittman\altaffilmark{1,3},
Georgiana Ogrean\altaffilmark{4,5},
Reinout van Weeren\altaffilmark{4},
Annalisa Bonafede\altaffilmark{6}}

\altaffiltext{1}{University of California, One Shields Avenue, Davis, CA 95616, USA}
\altaffiltext{2}{Lawrence Livermore National Laboratory, 7000 East Avenue, Livermore, CA 94550, USA}
\altaffiltext{3}{Instituto de Astrof\'{\i}sica e Ci\^{e}ncias do Espa\c{c}o, Universidade de Lisboa, Lisbon, Portugal}
\altaffiltext{4}{Harvard-Smithsonian Center for Astrophysics, 60 Garden Street, Cambridge, MA 02138, USA}
\altaffiltext{5}{Kavli Institute for Particle Astrophysics and Cosmology, Stanford University, 452 Lomita Mall, Stanford, CA 94305, USA}
\altaffiltext{6}{Hamburger Sternwarte, Universit\"at Hamburg, Gojenbergsweg 112, 21029 Hamburg, Germany}

\email{nrgolovich@ucdavis.edu}

\shortauthors{Golovich et al.}

\label{firstpage}

\begin{abstract}
We present an analysis of the merging cluster MACS J1149.5+2223 using archival imaging from Subaru/Suprime-Cam and multi-object spectroscopy from Keck/DEIMOS and Gemini/GMOS. We employ two and three dimensional substructure tests and determine that MACS J1149.5+2223 is composed of two separate mergers between three subclusters occurring $\sim$1 Gyr apart. The primary merger gives rise to elongated X-ray morphology and a radio relic in the southeast. The brightest cluster galaxy is a member of the northern subcluster of the primary merger. This subcluster is very massive (16.7$^{+\text{1.25}}_{-\text{1.60}}\times\text{10}^{\text{14}}$ M$_{\odot}$). The southern subcluster is also very massive (10.8$^{+\text{3.37}}_{-\text{3.54}}\times\text{10}^{\text{14}}$ M$_{\odot}$), yet it lacks an associated X-ray surface brightness peak, and it has been unidentified previously despite the detailed study of this \emph{Frontier Field} cluster. A secondary merger is occurring in the north along the line of sight with a third, less massive, subcluster (1.20$^{+\text{0.19}}_{-\text{0.34}}\times\text{10}^{\text{14}}$ M$_{\odot}$). We perform a Monte Carlo dynamical analysis on the main merger and estimate a collision speed at pericenter of 2770$^{+\text{610}}_{-\text{310}}$ km s$^{-\text{1}}$. We show the merger to be returning from apocenter with core passage occurring 1.16$^{+\text{0.50}}_{-\text{0.25}}$ Gyr before the observed state. We identify the line of sight merging subcluster in a strong lensing analysis in the literature and show that it is likely bound to MACS J1149 despite having reached an extreme collision velocity of $\sim$4000 km s$^{-\text{1}}$. 
\end{abstract}


\section{Introduction}\label{sec:intro}
Galaxy clusters grow continually through accretion of matter from the large scale filaments of the cosmic web. The largest of these clusters lie at the nodes, and accretion and relaxation occur continually over a wide range of mass scales \citep{EvrardReview}. Occasionally, clusters as massive as $\sim\text{10}^{\text{15}}\,\text{M}_{\odot}$ collide in major mergers. These events provide a key laboratory for understanding the physics of the main components: dark matter (DM), the hot intra-cluster medium (ICM), and the galaxies. In many of the most violent collisions, the ICM is stripped from the effectively collisionless DM and galaxies \citep[e.g. the Bullet Cluster and the Sausage Cluster;][]{Markevitch04, Clowe06,Dawson:2014, Jee:2014, Stroe:2014b, Sobral:2015}. These mergers are said to be dissociative \citep{Dawson:2012}.

Many mergers have proven to be more complex than the Bullet cluster and the Sausage cluster. The HST Frontier Field \citep{Lotz2014} clusters in particular appear to host multiple merging events rather than two-component, head-on collisions. These systems provide the community with a substantial challenge. Understanding many of the interesting phenomena requires an accurate quantification of the dynamics of the merging event. For example, DM-gas and DM-galaxy offsets have the potential to constrain the self-interacting cross section of DM \citep{Markevitch04, Clowe06, randall2008, Merten:2011}, but in order to translate a measured offset to a cross section, we must understand the dynamics of the merger (age, speed at collision, and viewing angle at the very least). \cite{Dawson:2012} demonstrated a modern version of the timing argument for bimodal mergers, but the parameter space is large and difficult to fully explore with simulations. Not all mergers are head on and bimodal, so it is an important step to be able to interpret more complex systems. 

\begin{figure*}[!htb]
\begin{center}
\includegraphics[width=\textwidth]{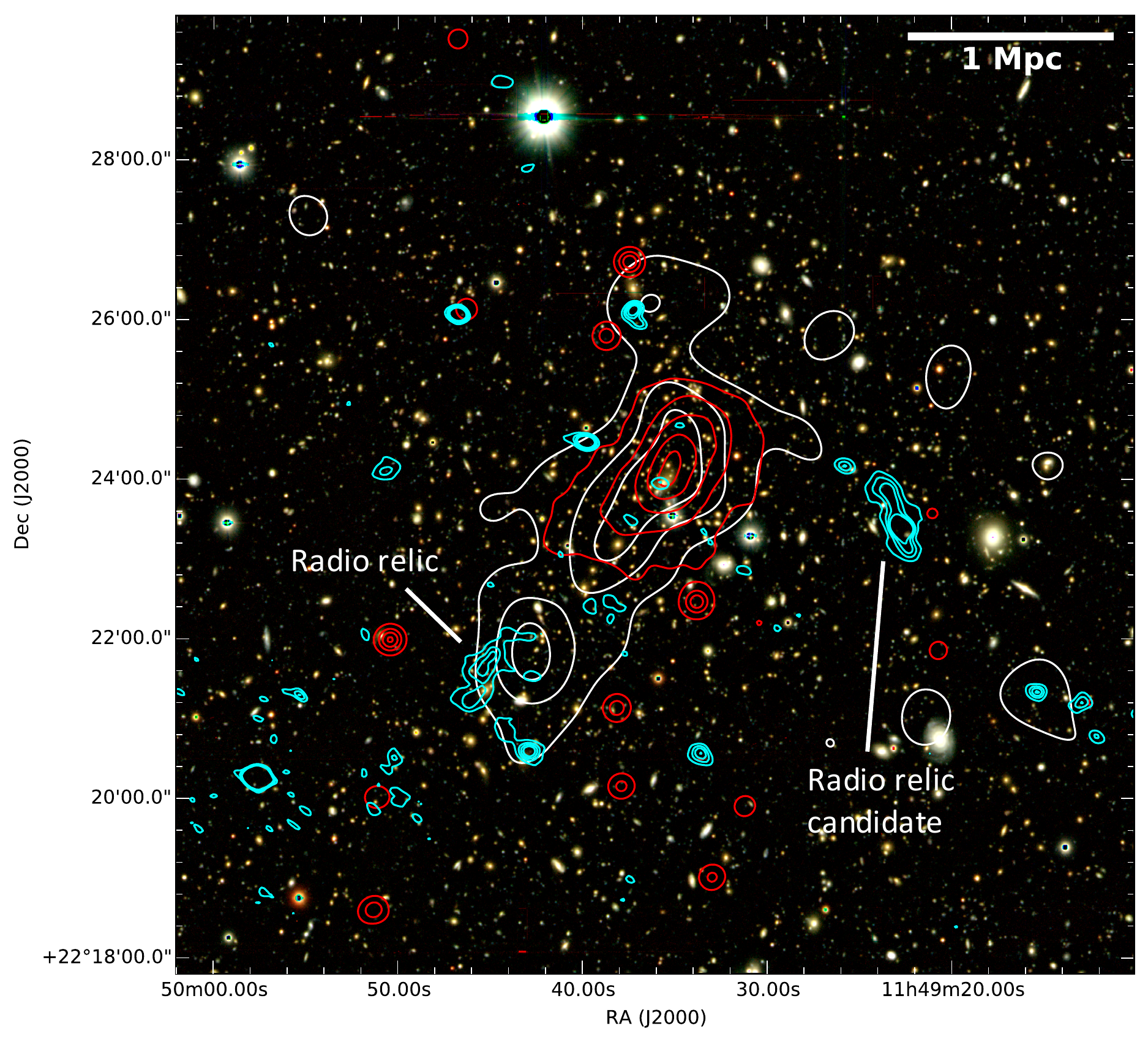}
\caption{Japanese Virtual Observatory Subaru VRi color composite image of MACS J1149. The red contours represent the Chandra X-ray surface brightness map (scaled linearly) presented by \cite{Ogrean:2016}. The elongation and extent of the X-ray surface brightness is indicative of a merger along a northwest--southeast axis. The white contours are a linear scale mapping of the red sequence galaxy R-band luminosity density, which is dominated by bright cluster members aligned along the same axis as the ICM. Interestingly, the southern galaxy subcluster lacks an associated X-ray surface brightness peak and is previously unidentified in the literature. The cyan contours are a linear scale mapping of GMRT 325 Mhz data presented in \cite{Bonafede12}. The two radio relics are the two large extended radio emission features. One of the relics lies to the southeast coincident with the supposed merger axis. The other lies to the west of this axis and is considered a relic candidate.}
\label{fig:rgb}
\end{center}
\end{figure*}

MACS J1149.5+2223 (hereafter MACS J1149) is a Frontier Field cluster at a redshift of 0.544 with a wealth of data from across the electromagnetic spectrum. It is an excellent system to test our ability to reassemble and interpret a complicated merger scenario (see Figure \ref{fig:rgb}).

\cite{Ebeling:2007} were the first to report on MACS J1149 as part of the MAssive Cluster Survey \citep[MACS;][]{Ebeling:2001}. They presented a wealth of information for the system including BIMA, Chandra/ACIS-I, Subaru/SuprimeCam BVRiz$\prime$, CFHT/MegaCam U, and HST/ACS F555W and F814W. They followed up with Keck/DEIMOS and Gemini/GMOS multi-object spectrometry, which were presented in \cite{Ebeling:2014}. Of the MACS clusters, MACS J1149 has the highest reported velocity dispersion of $\sim\text{1800}\,\text{km s}^{-\text{1}}$ as well as an X-ray luminosity of $\sim\text{18}\times\text{10}^{\text{44}}\,\text{erg}\,\text{s}^{-\text{1}}$ and temperature of $\sim\,\text{9}\,\text{keV}$ \citep{Ebeling:2007}. With this strong indication that MACS J1149 was a very massive cluster, \cite{Smith09} followed with a strong lensing analysis of the system based on HST/ACS imaging and Keck spectroscopic confirmation of multiply imaged galaxies finding a mass within 500 kpc of 6.7$\pm$0.4$\times$10$^{\text{14}}$M$_{\odot}$. They found the core to be composed of a dominant structure associated with the BCG along with three additional lensing halos situated along a northwest-southeast axis. \cite{Zitrin09} published an independent strong lensing analysis of the system and noted the largest known lensed image of a single spiral galaxy. This galaxy is a face-on spiral at a redshift of 1.49 and has been multiply imaged by MACS J1149. One of these images contained a supernova, now known as SN Refsdal, lensed into an Einstein cross pattern. The supernova should appear in another of the images in the future and the time difference between the appearance of the supernova can be used to constrain cosmology and the lens model of MACS J1149 \citep{Refsdal:1964, Kelly:2015, Sharon:2015, Oguri:2015, Diego:2016, Treu:2016}. \cite{Zitrin09} argue that a nearly uniform mass distribution over a 200 kpc radius with a surface density near the critical density is needed to explain the small amount of image distortion in this spiral. They further note that this mass distribution served as the most powerful cosmic lens known at the time. As evidence of the tremendous lensing potential, \cite{Zheng2012} used this powerful lens magnification to discover a $\text{z}\sim\text{10}$ galaxy. These results contributed to MACS J1149 being selected as part of the Cluster Lensing and Supernova Survey \citep[CLASH;][]{Postman:2012}. As part of this survey, MACS J1149 was studied with ground based weak lensing with Subaru/SuprimeCam \citep{Umetsu:2014}, and with a joint weak and strong lensing analysis utilizing Subaru/SuprimeCam and HST/ACS \citep{Umetsu:2015}, respectively. Both analyses resulted in similar a value for M$_{\text{200}}$: 25.4$\pm$5.2$\times$10$^{\text{14}}$ M$_{\odot}$ and 25.02$\pm$5.53$\times$10$^{\text{14}}$ M$_{\odot}$, respectively making MACS J1149 among the most massive clusters \citep{PlanckMass}. 

Diffuse radio sources, such as radio relics and radio halos, are often associated with major cluster mergers \citep{Feretti:2012}. Using GMRT and VLA radio observations, \cite{Bonafede12} reported evidence for three such sources in MACS J1149: a radio relic southeast and a candidate relic west of the cluster center as well as a possible radio halo located near the peak of the X-ray gas distribution. By combining archival 1.4 GHz Very Large Array (VLA) data with 323 MHz Giant Metrewave Radio Telescope (GMRT) data they suggested that the radio halo has an extremely steep spectral index of $\alpha \approx \text{2}$ indicating the cluster merger is not recent and/or the (re)acceleration process was not efficient \citep{Brunetti:2008}. While it was noted that the relics' major axes were oriented tangentially to the cluster center (see Figure \ref{fig:rgb}), the two relics are not symmetrically situated about the cluster center. If the merger axis is inferred from the gas distribution (northwest--southeast orientation), only the southeast relic is collinear. The other relic is located to the west, nearly perpendicular to the merger axis. \cite{Bonafede12} noted that this is uncommon for double radio relic systems and suggest that this may be an indication of multiple ongoing mergers in the system. The western relic appears near both a radio and X-ray point source, so it is possible that it is not a relic at all. For these reasons, \cite{Bonafede12} classified the west radio feature as a candidate. 

\cite{Ogrean:2016} present a deep (365 ks) Chandra X-ray analysis of the cluster. They report MACS J1149 to be among the most X-ray luminous ($L_{X,\text{[0.1--2.4\,keV]}}=\text{(1.62}\pm\text{0.02)}\times\text{10}^{\text{45}}\,\text{erg}\,\text{s}^{-\text{1}}$) and hottest ($T_{X} = \text{10.73}^{+\text{0.62}}_{-\text{0.43}}\,\text{keV}$) clusters measured to date. Despite the clear merger activity in the cluster, the X-ray surface brightness appears to be relatively regular for a multi-component merger indicating that MACS J1149 could be an old merger. This agrees with the steep spectral index of the radio halo. However, they detect a distant cold front to the northeast of the cluster as well as a possible surface brightness edge that could be the leading bow shock, which could indicate that merger activity is still present or that sloshing may be occurring in the ICM. 
 
In this paper, we extend the understanding of MACS J1149 with an optical and spectroscopic galaxy substructure analysis as well as an analysis of the merger dynamics. This paper is organized as follows. In \S\ref{sec:data}, we highlight the dataset we will be examining, and we describe our galaxy cluster membership selection. In \S\ref{sec:subclusters}, we describe our subcluster analysis. In \S\ref{sec:scenario}, we interpret the merger scenario and discuss our findings in light of the present understanding of this cluster. In \S\ref{sec:mcmac} we analyze the merger dynamics of the two merger events, and in \S\ref{sec:discussion}, we discuss, summarize, and present our conclusions. We adopt a flat $\Lambda$CDM universe with $H_0 = 70$ $\text{km}$ $\text{s}^{-1}$ $\text{Mpc}^{-1}$, $\Omega_M = 0.3$, and $\Omega_{\Lambda} = 0.7$. At the cluster redshift ($z = 0.544$), 1$\arcmin$ corresponds to 383 kpc.


\section{Data acquisition and preparation}\label{sec:data}

Our primary goal is to understand the merger dynamics. To do so, we make use of two sets of archival data: 
\begin{itemize}
\item Photometric catalog from the CLASH survey \citep{Umetsu:2014} with data from Subaru/SuprimeCam
\item Spectroscopic catalog from \cite{Ebeling:2014} with data from Keck/DEIMOS and Gemini/GMOS \citep[see Table 2 of][]{Ebeling:2014}
\end{itemize}
While the photometric catalog has the advantage of being much more complete in the region of interest, it suffers from contamination by non-cluster-member objects. Meanwhile, the spectroscopic catalog has high purity and gives valuable insight into line of sight (LOS) information, but it suffers from selection effects and is incomplete. 

\begin{figure}[!htb]
\includegraphics[width=\columnwidth]{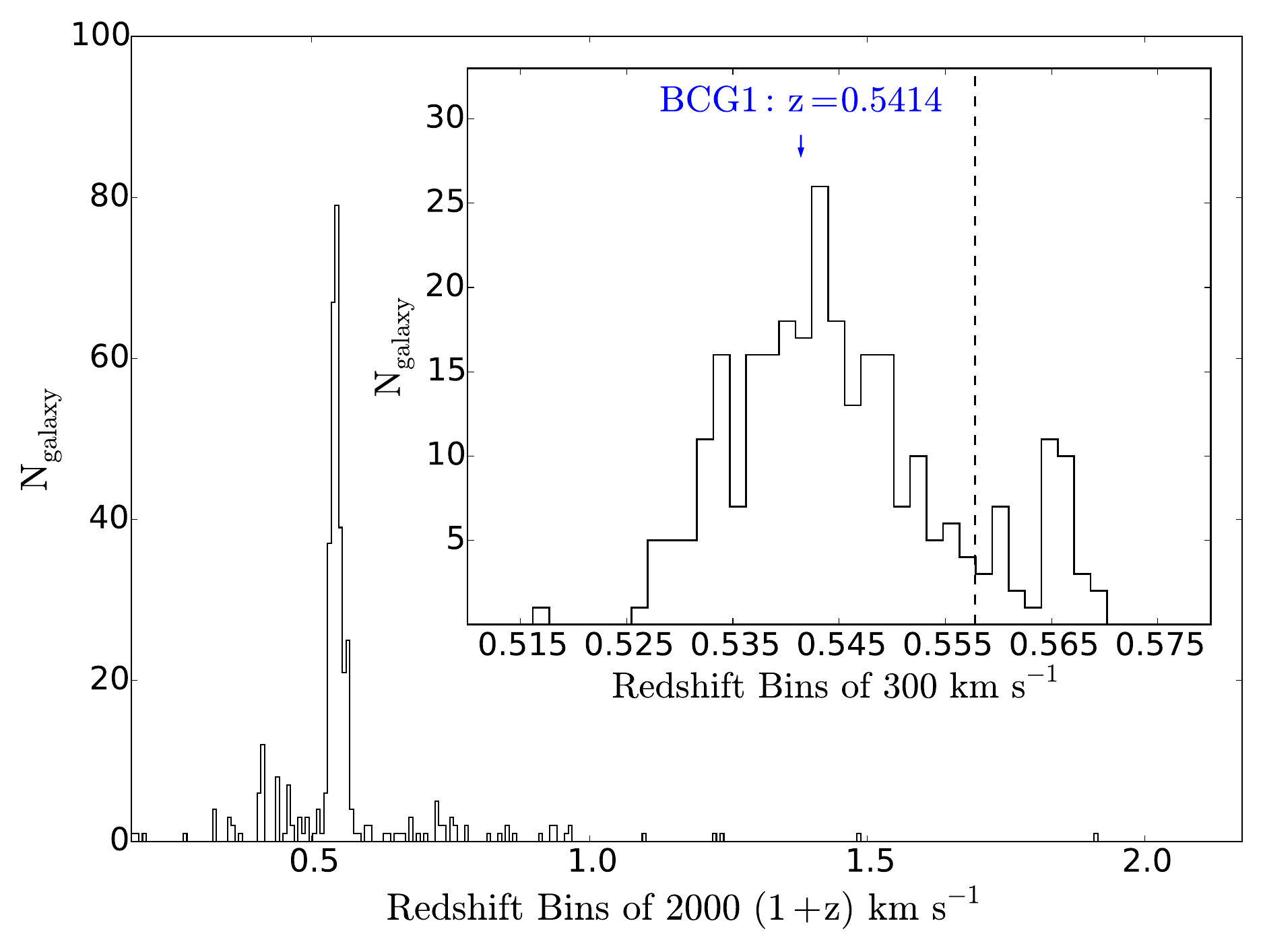}
\caption{\emph{Main:} Redshift distribution for 391 spectroscopic redshifts in the 5x5 Mpc MACS J1149 field. These galaxies are a subset of the original spectroscopic catalog presented by \cite{Ebeling:2014}. We discuss the cluster member selection in \S\ref{sec:data}. \emph{Inset:} A zoomed view of the 278 spectroscopically confirmed cluster members of MACS J1149 with the BCG identified by its redshift. The dashed line is at a redshift of 0.558, and the analysis in \S \ref{sssec:dstest} will study the substructure above and below this redshift both together and separately.\\} 
\label{fig:hist}
\end{figure}

\begin{figure}[!htb]
\includegraphics[width=\columnwidth]{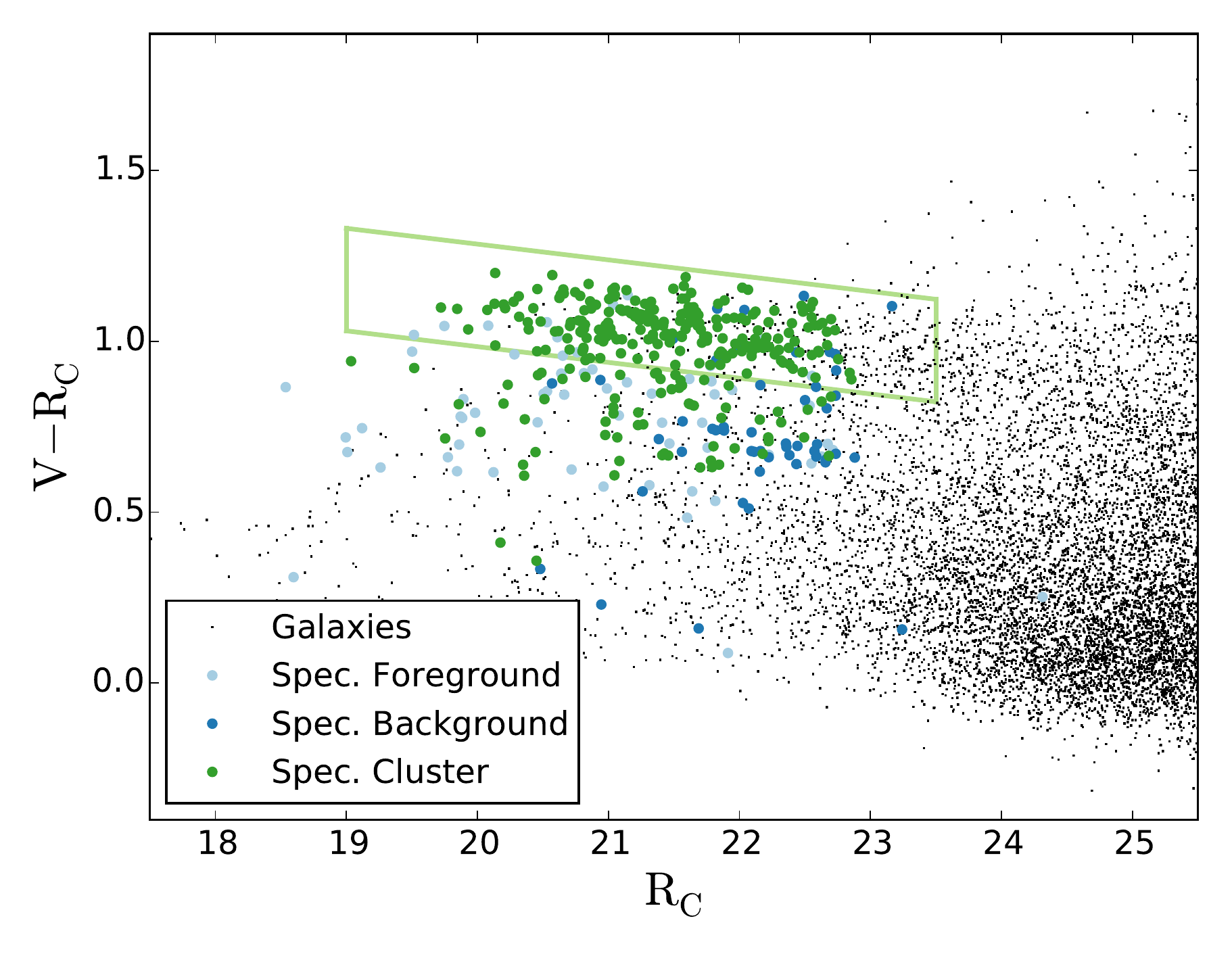}
\caption{Color-magnitude diagram of galaxies within a square 5x5 Mpc overlapping the spectroscopic survey based on Subaru V and R$_{\text{C}}$ magnitudes. Spectroscopically confirmed cluster (green), foreground (light blue), and background (dark blue) galaxies are overlaid. Our red sequence selection region is outlined in light green.}
\label{fig:redsequence}
\end{figure}

We restrict both catalogs to a 5x5 Mpc region about the center of the cluster. The photometric catalog has 14,739 objects (87 arcmin$^{-\text{2}}$) and the spectroscopic catalog has 391 objects. To aid in our red sequence selection, we identify cluster members as any galaxy with spectroscopic redshift between $\pm\text{5400}\,\text{km}\,\text{s}^{-\text{1}}$ of the average redshift of $\text{0.544}$, which corresponds to a range in redshift of $\text{0.516255}<\text{z}<\text{0.571745}$. The velocity window is chosen based on the velocity dispersion reported by \cite{Ebeling:2014}. Within these limits, there are 278 spectroscopic cluster members, and outside there are 62 foreground galaxies and 51 background galaxies in the spectroscopic catalog. We present these redshifts in Figure \ref{fig:hist}. The CLASH team already removed the stars from the photometric catalog, so our task is to separate cluster members from foreground and background galaxies. We first identify the red sequence of the cluster, which we highlight with spectroscopically confirmed cluster members by matching the two catalogs with the Topcat \citep{Topcat} software using a 1$\arcsec$ tolerance. We overlaid the spectroscopically confirmed objects with the rest of the photometric catalog on a color-magnitude diagram, where we also define the red sequence (Figure \ref{fig:redsequence}). There are 542 total galaxies within our red sequence selection box. Of these, 233 galaxies have spectroscopic redshifts (209 cluster members, 12 background galaxies, 12 foreground galaxies). The spectroscopic survey targeted cluster members, so it is not possible to fully rule out foreground and background clusters. \cite{Ogrean:2016} pointed out Abell 1388 ($z=0.18$) sitting approximately 5$\arcmin$ to the west. We also identified a small foreground group at a redshift of 0.42 sitting 1.5$\arcmin$ to the east. Our red sequence selection eliminated galaxies at these redshifts from our photometric catalog.

\section{Subcluster analysis}\label{sec:subclusters}

In the following subsections, we will make use of the spectroscopic catalog for a one dimensional subcluster analysis (\S\ref{subsec:oned}), the red sequence catalog for a projected, two dimensional analysis (\S\ref{subsec:twod}), and we use the spectroscopic catalog for two separate three dimensional analyses (\S\ref{subsec:threed}). 

\subsection{One dimensional analysis}\label{subsec:oned}

Before these more detailed analyses are discussed, we will inspect the velocity distribution of the spectroscopic catalog presented in Figure \ref{fig:hist}. The cluster is clearly composed of a dominant velocity component near $z=0.544$; however, there is a separate higher redshift group of galaxies. A Kolmogorov-Smirnov test for Gaussianity results in a p-value of 0.052, which suggests tension with the null hypothesis of a single Gaussian component. When we removed the galaxies above $z=0.558$ (dashed line in Figure \ref{fig:hist}), the p-value is 0.88 strongly supporting a single Gaussian distribution of redshifts. However, this test is insensitive to substructure near the same redshift, so we also conduct two and three dimensional analyses in the following subsections. 

\subsection{Two dimensional analysis}\label{subsec:twod}

We generate a red sequence projected galaxy luminosity distribution by weighting galaxy locations by their observed R-band luminosity assuming each galaxy to be at the average distance of the cluster. We smooth the map with a 20$\arcsec$ Gaussian kernel and plot contours. The density map is presented as contours in Figure \ref{fig:rgb} and also in grayscale with linearly separated contours in Figure \ref{fig:dstest}.

The red sequence galaxy distribution displays an elongated morphology in the same direction as the X-ray surface brightness; however, there is clear bimodality instead of a single surface brightness concentration. A second subcluster now appears without a corresponding X-ray concentration as evident in Figure \ref{fig:rgb}. This subcluster has not been previously identified in the literature. The X-ray emission appears to be associated with the northern subcluster, which corresponds to the region of MACS J1149 that has been studied previously in the literature and contains the BCG. We located the peak of the luminosity weighted density map and performed a bootstrap-uncertainty analysis with our red sequence catalog to find the uncertainty in the peak locations of our density map. The northwestern peak is located at RA$=$11$^{h}$49$^{m}$36.81$^{+\text{4.3}}_{-\text{8.5}}$$^{s}$, DEC$=$22$^{\circ}$23$\arcmin$57.2$^{+\text{2.4}}_{-\text{6.1}}\arcsec$ and the southeastern peak is located at RA$=$11$^{h}$49$^{m}$42.9$^{+\text{4.3}}_{-\text{12.7}}$$^{s}$, DEC$=$22$^{\circ}$21$\arcmin$56.0$^{+\text{13.3}}_{-\text{9.7}}\arcsec$. We find the projected separation between the two to be 2.60$^{+\text{0.32}}_{-\text{0.40}}\arcmin$ (0.99$^{+\text{0.12}}_{-\text{0.15}}$ Mpc). The red sequence luminosity map shows MACS J1149 to be a bimodal system, with the northwest peak more extended and luminous than the southeast peak. It has been said that MACS J1149 is likely composed of multiple mergers \citep{Smith09, Bonafede12}. This is not evident from the red sequence luminosity distribution alone.

\subsection{Three dimensional analyses}\label{subsec:threed}

We employ two methods of subcluster identification with the spectroscopic catalog. The first is the Dressler-Shectman test \citep[DS-test;][]{Dressler:1988} and the second is a Markov Chain Monte Carlo (MCMC) Gaussian Mixture Model (GMM) analysis, which we will refer to as MCMC-GMM. The DS-test will highlight substructure, but it is incapable of making quantitative estimates of subcluster velocity information or the likelihood of subcluster models. As such, the results from the DS-test and the red sequence luminosity density map are used in conjunction with existing data from the literature to provide informative priors for our MCMC-GMM analysis.

\subsubsection{Dressler Schectman Test} \label{sssec:dstest}

The Dressler-Schectman (DS)-test is performed by computing a $\chi^{2}$-like statistic for local velocity information as compared to the global values for each galaxy. The statistic is given by:
\begin{equation}\label{eq:delta}
\delta^{2} = \frac{N_{local}}{\sigma^{2}}\big[\left(\bar{v}_{local}-\bar{v}\right)^{2} + \left(\sigma_{local} - \sigma\right)^{2}\big]
\end{equation}
where $N_{local}$ is the number of nearest neighbors (self-inclusive) to include when calculating $\bar{v}_{local}$, the local-average LOS velocity, and $\sigma_{local}$, the local velocity dispersion. We let $N_{local} \equiv \lceil\sqrt{N_{total}} \rceil$, where $N_{total}$ is the number of galaxies in the full spectroscopic catalog. This follows the best practice identified by \cite{Pinkney:1996}. Galaxies with larger $\delta$ values are highly correlated with their neighbors and different from the parameters thus identifying local structure.

\begin{figure*}[!htb]
\begin{center}
\includegraphics[width=\textwidth]{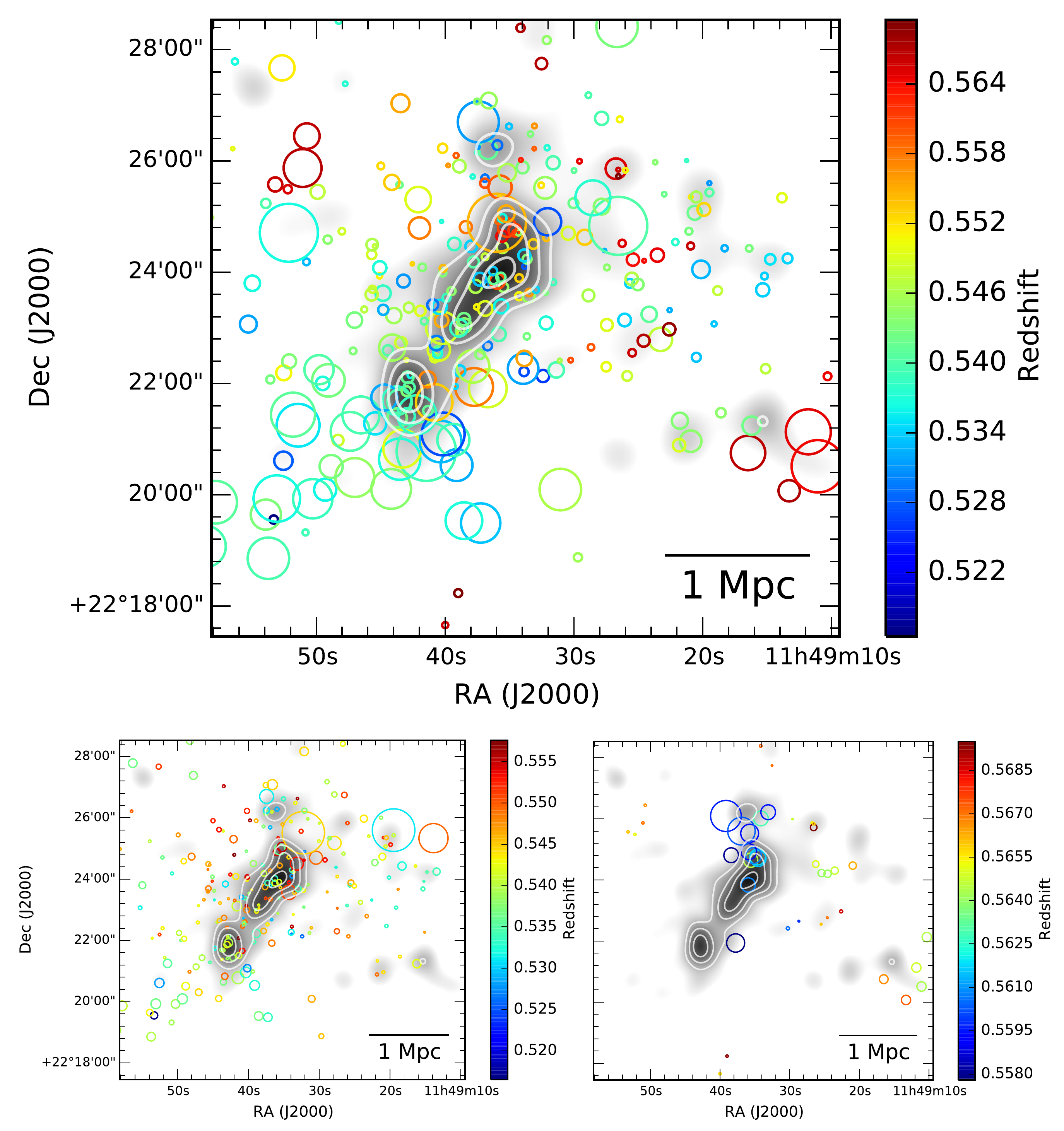}
\caption{\emph{Top:} DS-test plot for MACS J1149 with projected locations of 278 spectroscopic cluster members color coded according to their redshift. The grayscale and linearly spaced contours represent the red sequence luminosity density. The diameter of each circle is proportional to $5^{\delta}$, where $\delta$ is the DS-$\delta$ given by Eq.~\ref{eq:delta}. Collections of larger than average circles trace substructure. The color of each circle corresponds to the redshift of the galaxy. \emph{Bottom left:} Same plot as above but only including galaxies below $z=0.558$. \emph{Bottom right:} Same plot as above but only including galaxies above $z=0.558$.}
\label{fig:dstest}
\end{center}
\end{figure*}

In Figure \ref{fig:dstest}, we plot the projected location of the galaxies in the spectroscopic catalog as circles with radii proportional to $5^{\delta}$ over the red sequence luminosity density map and linearly space contours. The top panel shows all galaxies in the spectroscopic catalog. The bottom left (right) panel shows galaxies below (above) $z=0.558$ (see Figure \ref{fig:hist}). Groups of enlarged circles imply substructure. We color-coded the circles in Figure \ref{fig:dstest} by redshift to aid the eye. The lack of significant structure in the lower left panel of Figure \ref{fig:dstest} suggests that the two main subclusters evident in the red sequence luminosity map are at a similar redshift. Both central regions appear to have an excess of yellow/green circles, which corresponds to a redshift slightly below the cluster average of 0.544. The two subclusters lie collinearly with the radio relic in the southeast as well as the extension of the ICM (see Figure \ref{fig:rgb}). These features suggest a post merger scenario and the similar colors of the two subclusters imply a merger axis near the plane of the sky. The northern structure is associated with the massive subcluster identified via strong gravitational lensing \citep{Smith09}. In the bottom right panel of Figure \ref{fig:dstest}, there is a clear over-density of galaxies near $z=0.56$ to the north of the main cluster. These galaxies correspond to the small spike of galaxies at $z=0.56$ in Figure \ref{fig:hist}. The group has a bright core that is also identified in the lensing analysis of \cite{Smith09} (labeled halo D). Additionally, there is a wall of higher redshift galaxies from west of the cluster to north of the cluster. These galaxies correspond to the over density of galaxies at $z=0.564$ in Figure \ref{fig:hist}. Both of these structures are at a higher redshift than the bulk of MACS J1149. There are relatively few galaxies at this redshift, which could either indicate that the structures are not very massive, or they are under sampled. We will discuss two possible scenarios in \S \ref{sec:scenario} and \S \ref{sec:mcmac}. First, the group is in the background of MACS J1149 along with the wall. Second, the group is not associated with the wall and is currently merging with the main cluster with an extremely high line of sight relative velocity of 3350$\pm$170 km s$^{-\text{1}}$, which would require it to be near pericenter and merging along the line of sight in the observed state. 

To measure the level of substructure statistically, we perform a bootstrap analysis in which we compute a cumulative structure parameter $\Delta\equiv\sum\delta_{i}$. We then randomly shuffle the redshifts amongst the projected locations of the spectroscopic cluster members computing $\Delta$ each time. For a system with little substructure, we expect $\Delta \sim N_{total}$. We completed this analysis for each of the three groups of spectroscopic redshifts (all spectra and those above and below a redshift of 0.558). The average value in the bootstrap sample for the full redshift catalog is $\Delta = 287 = 1.03 \times N_{total}$. For these data, $\Delta = 449 = 1.56 \times N_{total}$. Comparing this to the bootstrap sample, the data exhibit substructure at a 5.3$\sigma$ level indicating substantial clustering in the full spectroscopic catalog. For the lower redshift galaxies, the average value in the bootstrap sample is $\Delta = 244 = 1.03 \times N_{total}$. For the data, $\Delta = 266 = 1.12 \times N_{total}$. Comparing this to the bootstrap sample, the data exhibit substructure at a 0.8$\sigma$ level. This indicates that the main redshift distribution ($z<\text{0.558}$) does not show substantial subclustering. For the higher redshift galaxies, the average value in the bootstrap sample is $\Delta = 36.8 = 0.92 \times N_{total}$. For the spectroscopic galaxies above $z=0.558$, $\Delta = 71.9 = 1.80 \times N_{total}$. Comparing this to the bootstrap sample, the data exhibit substructure at a 4.1$\sigma$ level indicating that the galaxies above a redshift of 0.558 are substantially subclustered. 

In summary, we have identified two main subclusters that are also clearly evident in the red sequence luminosity map. These subclusters are approximately at the same redshift and are composed of galaxies making up the main over density of galaxies (below $z=0.588$) in Figure \ref{fig:hist}. The northwest subcluster is more extended than the southeast subcluster. A third line of sight structure is identified from the DS-test. It may have been recently decoupled from a wall of higher redshift galaxies in the background, or it may be involved in an extremely high-velocity merger with the northwest subcluster, we will discuss both of these scenarios quantitatively in \S\ref{sec:scenario} and \S\ref{sec:mcmac}. These galaxies appear in the redshift distribution to the right of the dashed line at  $z=0.558$ (see Figure \ref{fig:hist}). We will fit these potential subclusters to the spectroscopic catalog with Gaussian mixture models of varying complexity in the following section. 

\subsubsection{MCMC-GMM analysis} \label{sssec:MCMCGMM}

Guided by the candidate subcluster locations identified above, we seek to fit subcluster properties quantitatively. In this section we implement a Markov-chain Monte Carlo Gaussian mixture model analysis (MCMC-GMM). The basic goal is to simultaneously fit multiple Gaussians to the projected location and redshift distribution of a population of galaxies in a merging cluster environment. We utilize the python package \emph{emcee} \citep{emcee} for the MCMC sampling.

Based on the previous section, we select one to four multivariate Gaussians to be included in the mixture model. The means of the multivariate Gaussians ($\langle$RA$\rangle$, $\langle$DEC$\rangle$, $\langle$z$\rangle$) define the location while the scales of the multivariate Gaussians are defined by the projected size and velocity dispersion of each subcluster. This yields a 3$\times$3 covariance matrix with non-zero $\sigma_{\text{RA}}$, $\sigma_{\text{DEC}}$, $\sigma_{\text{z}}$, and cross term $\sigma_{\text{RA-DEC}}$. The other cross terms $\sigma_{\text{RA-z}}$ and $\sigma_{\text{DEC-z}}$ are zero because we do not expect rotation. Additionally, for each model, we include a background group, to account for the possibility of field galaxies. This background is modeled as a diagonal multivariate Gaussian with uninformative priors on its six parameters. 

We assume uniform priors on all other parameters. Specifically, we enforce the ratio of semi-minor to semi-major axes to be between 0.4 and 1 to avoid highly elliptical projections \citep{Schneider:2012}. We enforce the semi-major axis, as defined by the $\sigma$ of the projected Gaussian, to be between 0.25 and 1 Mpc to avoid overfitting or generating nonphysically large subclusters. The priors for the right ascension, declination, redshift, and velocity dispersion of each candidate subcluster are summarized in Table \ref{table:priors}. We make use of the lensing constraints of \cite{Smith09} for the velocity dispersion prior. We assume a uniform prior of width 3$\sigma$ based on the results in Table 1 of \cite{Smith09} for two subclusters: the primary subcluster in the north and the higher redshift group. The southern subcluster has not been discussed in the literature, but based on the red sequence luminosity distribution, the southeast subcluster appears to be of similar mass as the northern subcluster, so we implement a prior between 500 km s$^{-\text{1}}$ and 1300 km s$^{-\text{1}}$. For the wall-like structure, we keep the priors uninformative and allow the code to sample a large region of parameter space; however, we restrict the velocity based on the galaxies that appear to be members of this structure. 

\begin{table*}[!htb]
\begin{center}
\caption{Uniform prior ranges for the inputs of the four candidate subclusters of MACS J1149 for the MCMC-GMM analysis.}
\begin{tabular}{lllll}
Subcluster &	X (Mpc) &	Y (Mpc) &	z (1000 km s$^{\text{-1}}$) & $\sigma_{z}$ (1000 km s$^{\text{-1}}$) \\
\hline \\
1 &	-0.19, 0.19 &	-0.15, 0.33   &	-1.0, 0.25	& 1.1, 1.4\\
2 &	0.49, 0.69  &	-0.81, -0.57  &	-1.0, 0.25	& 0.5, 1.3\\
3 &	-0.30, 0.06 &	0.26, 0.59    &	2.7, 2.9	& 0.14, 0.78\\
Wall &	-3.0, 3.0 		&    -3.0, 3.0    	&    3.0, 5.0	& 0.2, 1.0\\
Background & -3.0, 3.0 & -3.0, 3.0 & -5.0, 5.0 & 2.0, 10\\
\end{tabular}
 \label{table:priors}
\end{center}
\end{table*}

\begin{table}[!htb]
\begin{center}
\caption{Subcluster membership and BIC scores for the four models run using MCMC-GMM}
\begin{tabular}{lcccccc}
Model			& N$_{1}$	& N$_{2}$	& N$_{3}$	& N$_{4}$	& N$_{field}$	& BIC-BIC$_{\text{min}}$\\
\hline \\
1			& 104	& -	& -	& -	& 174	& 91\\
2			& 106	& 56	& -	& -	& 116	& 11\\
3			& 133	& 59	& 14	& -	& 56		& 0\\
4			& 181	& 59	& 12	& 23	& 3		& 39\\
\end{tabular}
\label{table:models}
\end{center}
\end{table}

\begin{figure}[!htb]
\includegraphics[width=\columnwidth]{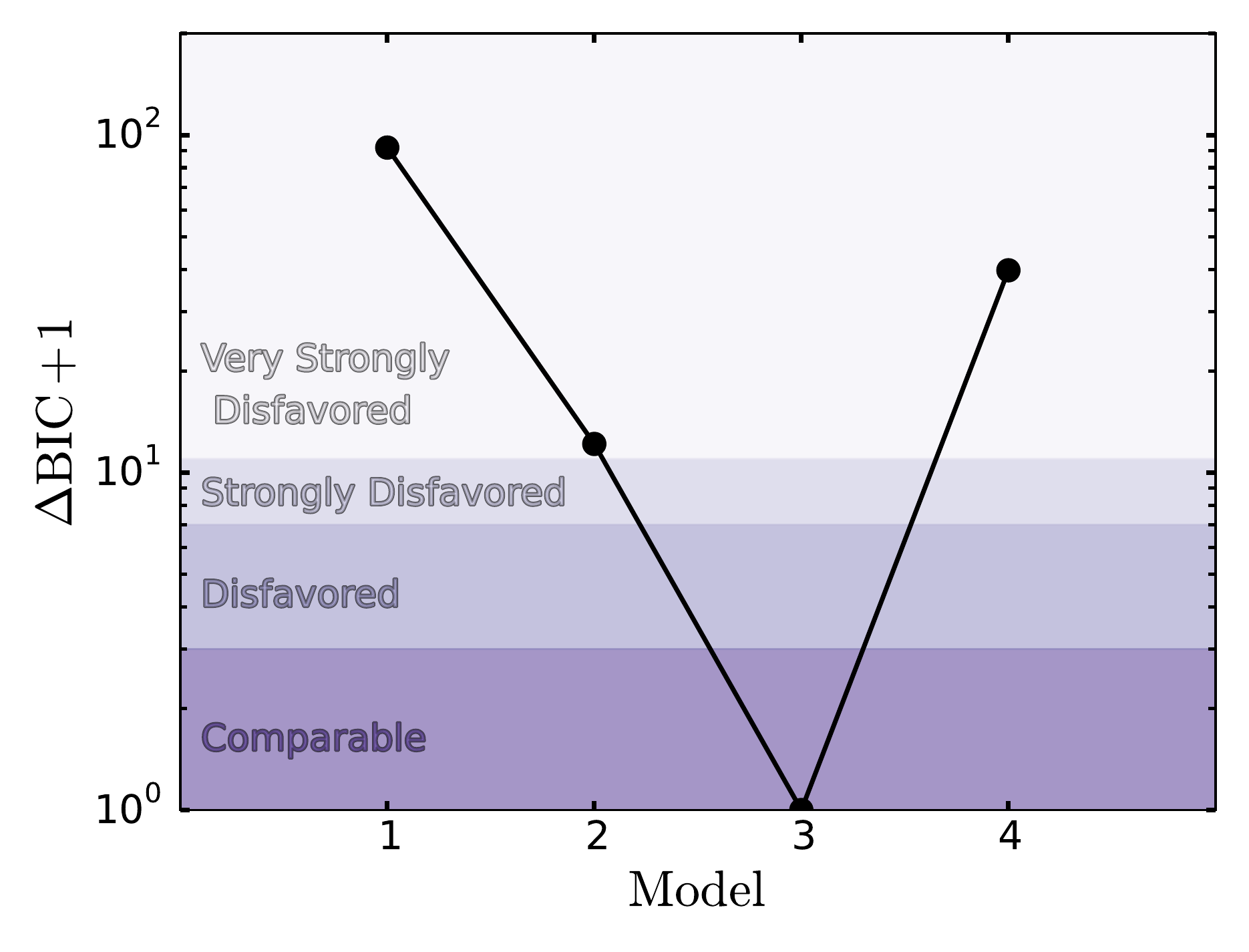}
\caption{$\Delta$BIC plot comparing MCMC-GMM fits to the three dimensional (right ascension, declination, and redshift) distribution of the spectroscopic cluster members determined in \S \ref{sec:data}. We plot the results for each model's BIC score relative to the lowest score achieved by the four models. The purple shaded regions roughly denote how a given model compares with the model that has the lowest score \citep{Kass:1995}. The best fit is model three (see Table \ref{table:models}).}
\label{fig:deltabic}
\end{figure}

\begin{figure*}[!htb]
\begin{center}
\includegraphics[width=\textwidth]{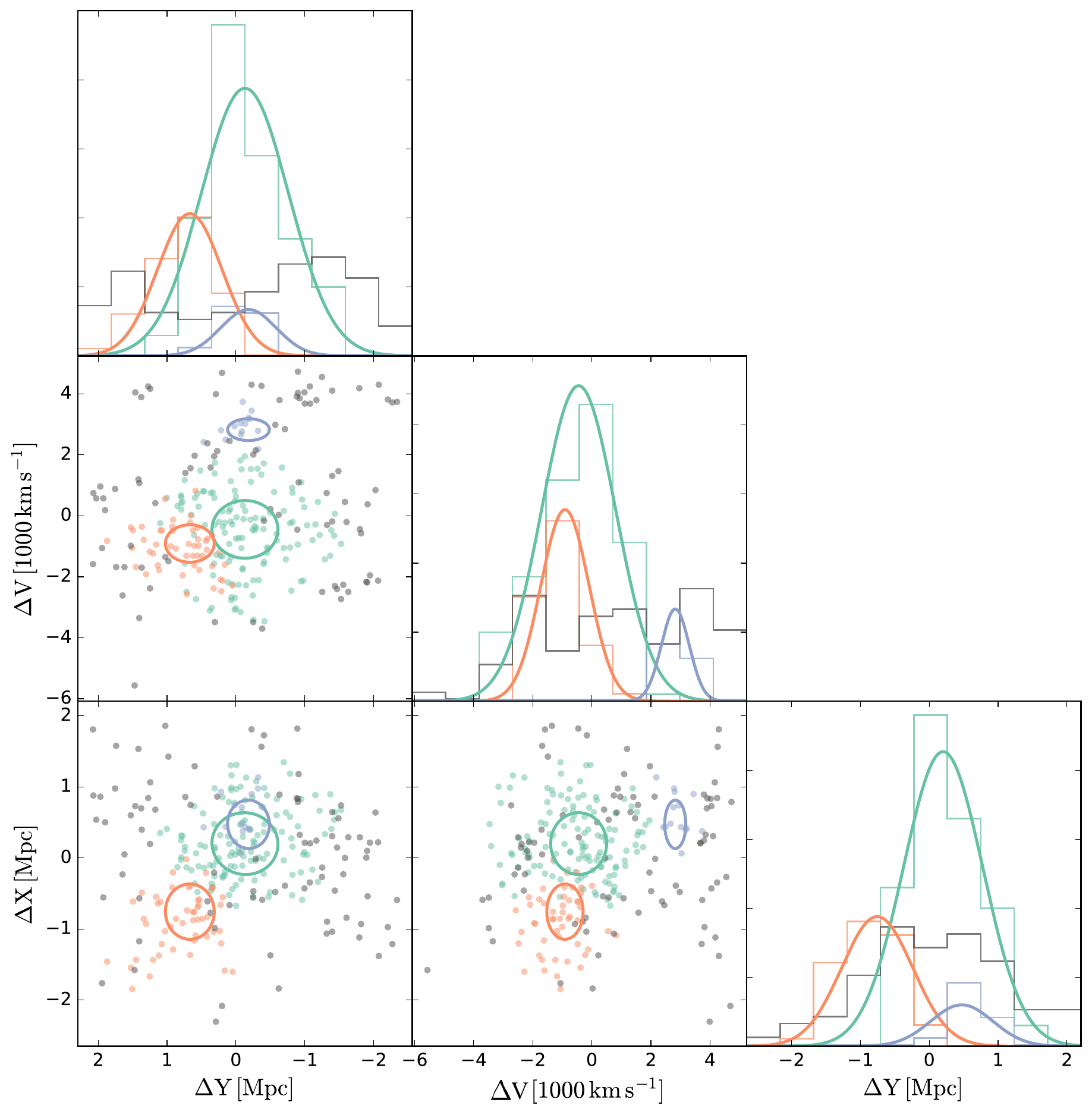}
\caption{Three dimensional corner plot for the subcluster membership for the 278 galaxies of the spectroscopic catalog (see \S\ref{sec:data}) determined by our MCMC-GMM analysis. Subclusters one, two, and three are labeled by green, orange, and blue points respectively. The gray points correspond to spectroscopically confirmed galaxies at the cluster redshift that are not included in any of the three subclusters. They should be interpreted as field galaxies. The X, Y, and z scales are measured with respect to the cluster averages from the spectroscopic catalog ($\langle$RA$\rangle=$11$^{h}$49$^{m}$36.7$^{s}$, $\langle$DEC$\rangle=$22$^{\circ}$23$\arcmin$40.3$\arcsec$, and $\langle$z$\rangle=$0.5452).}
\label{fig:trifit}
\end{center}
\end{figure*}

We run MCMC-GMM with 400 walkers each taking 10,000 steps burning the first 5,000 steps to allow for convergence before posteriors are used to infer parameters. We checked that the number of steps is more than ten times the autocorrelation time to ensure convergence. We run four types of models, and each model's Gaussians' galaxy membership is outlined in Table \ref{table:models}.  Each model has a background component. Model 1 through Model 4 corresponds to the number of subclusters (excluding the background component) to be fit. For each run, we select the realization that results in the maximum likelihood for fitting the data. We then compare the four models with a Bayesian Information Criterion (BIC), which measures the likelihood of each model while penalizing the models with more subclusters to avoid rewarding over-fitting. A lower BIC score indicates the data are more economically described by the model. We summarize the results of this analysis in Figure \ref{fig:deltabic}, in which we plot the difference in BIC scores relative to the lowest BIC score among the models. $\Delta$BIC$>$10 indicates a worse fit of a model \citep{Kass:1995}, so the models containing one, two, and four subclusters are rejected. We will focus our analysis on the three component model from here.

In Figure \ref{fig:trifit}, we plot a three dimensional corner plot for the preferred model, containing the color-coded membership selection for each spectroscopic galaxy. Galaxy membership is assigned to the subcluster with the highest likelihood of hosting each galaxy based on the three dimensional Gaussian for each subcluster. The subcluster Gaussians are presented as projected ellipses centered on the most likely value with widths corresponding to the marginal 1-$\sigma$ confidence in the X, Y, and z positions. The X, Y, and z axes are displayed as differences with respect to the cluster averages, which are listed in the caption. For the velocity axis, we use units of $\text{1000}$ km s$^{-\text{1}}$, which sets the scale to be of $\mathcal{O}(1)$ like the projected dimensions (in units of Mpc). Subclusters one and two (red and blue in Figure \ref{fig:trifit}, respectively) have the largest populations, and the higher redshift subcluster was assigned a compact group of 14 galaxies (green in Figure \ref{fig:trifit}). The locations of these subclusters are summarized in Table \ref{table:results}. Galaxies in gray are assigned to the background model. These include all of the galaxies that appear as a wall-like structure discussed above in \S\ref{sssec:dstest}. This makes sense given the large spatial spread in these galaxies. Our four component model was rejected by the BIC score, but it was able to properly classify these galaxies into the expected wall-like structure. We utilized a cluster based prior, so we would not expect a wall-like structure to be fit better by the model than the background model. These galaxies are not involved in the merger and thus are not of primary interest, so we will not discuss them further. 

\begin{table*}[!htb]
\begin{center}
\caption{Results for the three subclusters of MACS J1149 after membership selection with the MCMC-GMM code and biweight analysis described in \S\ref{sssec:MCMCGMM}}
\begin{tabular}{llllllll}
Subcluster &	N &	RA &	DEC &	Redshift &	$\sigma_{v}$ (km s$^{-1}$) &	$\sigma_{v}\,M_{200}$ ($10^{14}\,M_{\odot}$) &	v - $\bar{v}$ (km s$^{-1}$)  \\
\hline \\
1 	& 133 	& 11$^{h}$49$^{m}$35.6$\pm$4.5$^{s}$ 	& 22$^{\circ}$24$^{\prime}$11.0$\pm$6.0$^{\prime\prime}$	& 0.54256$^{+0.00085}_{-0.00077}$ & 1260$^{+31}_{-42}$	& 16.71$^{+1.25}_{-1.60}$ 	& -504$\pm$158 \\
2 	& 59		& 11$^{h}$49$^{m}$43.2$\pm$2.9$^{s}$ 	& 22$^{\circ}$21$^{\prime}$51.4$\pm$0.9$^{\prime\prime}$ 	& 0.54102$^{+0.00091}_{-0.00065}$ & 1088$^{+104}_{-136}$ & 10.80$^{+3.37}_{-3.54}$ 		& -806$\pm$152 \\
3	& 14 		& 11$^{h}$49$^{m}$35.5$\pm$6.5$^{s}$ 	& 22$^{\circ}$24$^{\prime}$49.5$\pm$1.6$^{\prime\prime}$ 	& 0.55988$^{+0.00026}_{-0.00039}$ & 522$^{+26}_{-55}$ 	&  1.20$^{+0.19}_{-0.34}$ 	&  2845$\pm$63 \\
\end{tabular}
\label{table:results}
\end{center}
\end{table*}

\begin{figure}[!htb]
\includegraphics[width=\columnwidth]{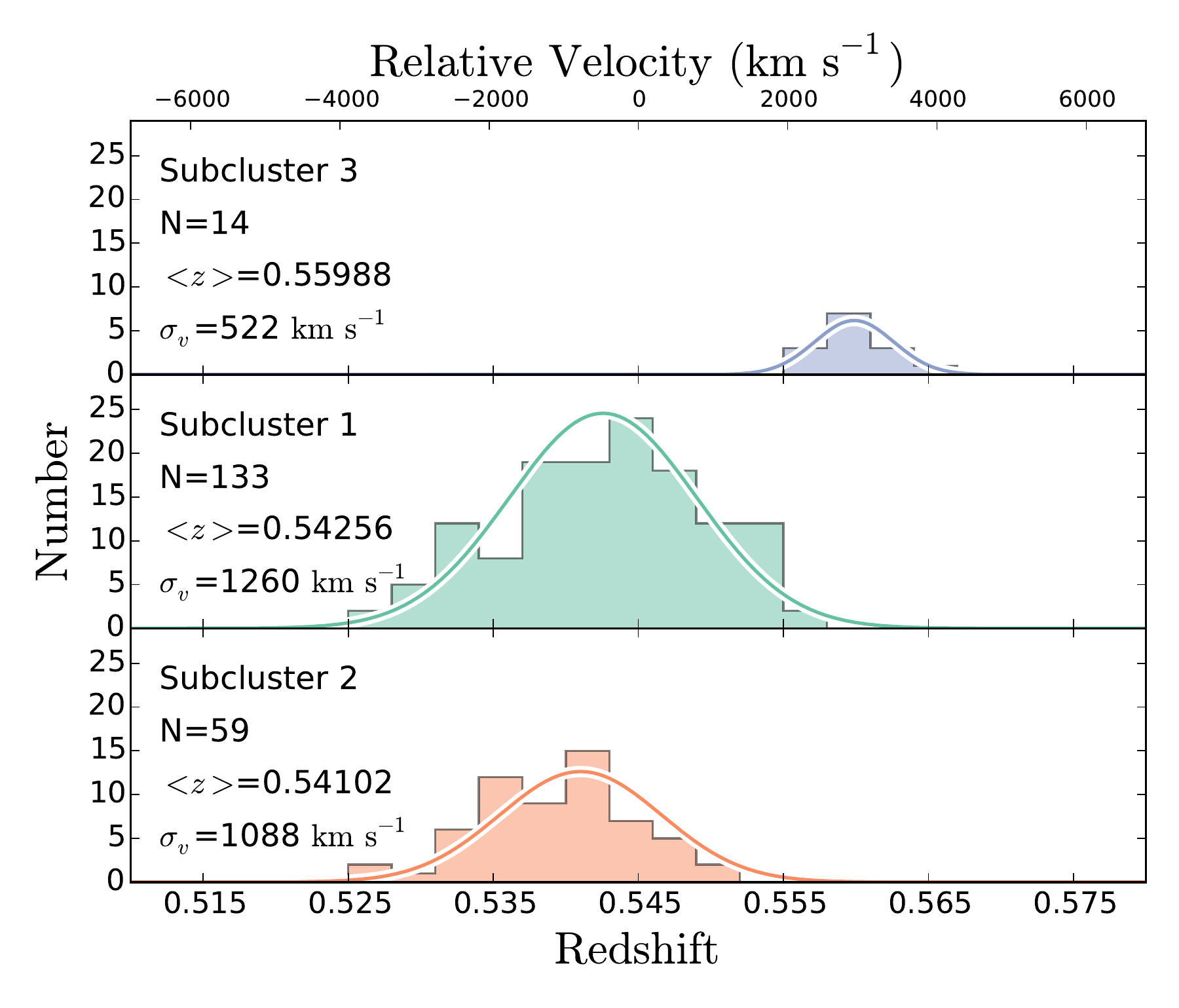}
\caption{Redshift distributions of the three subclusters selected by the MCMC-GMM code. The panels are arranged north to south and the colors match the colors of Figure \ref{fig:trifit}. The membership, biweight-analysis redshift and velocity dispersions (see Table \ref{table:results}) are displayed. A normalized Gaussian with these parameters is overlaid. The velocity scale at the top of the figure is relative to $z=0.545$.}
\label{fig:subclusthist}
\end{figure}

We implement the biweight statistic based on bootstrap samples of each subcluster's member galaxies and calculate the bias-corrected 68\% confidence limits for the redshift and velocity dispersion from the bootstrap sample. This method is more robust to outliers than the dispersion of the Gaussians generated by our statistical model \citep{Beers1990}. The redshift distributions of the subclusters are shown in Figure \ref{fig:subclusthist}. We use the velocity dispersions to estimate the M$_{\text{200}}$ for each subcluster using the \cite{Evrard:2008} scaling relations. We summarize these subcluster properties in Table \ref{table:results}. We use the biweight analysis to mitigate outliers from affecting the estimates of the velocity dispersion and redshift, but this does not protect against the well known fact that mergers tend to bias high the velocity dispersion based mass estimates \citep{Takizawa:2010, Saro:2013}. We discuss this in more detail in \S \ref{subsec:vdisp_mass}. 

\section{Merger scenario}\label{sec:scenario}

We identify three subclusters (subclusters one, two, and three in red, blue, and green respectively from \S\ref{sssec:MCMCGMM}). Subclusters one and two make up 96$\%$ of the mass and are post-merging subclusters along a northwest-southeast axis that generated the elongation of the X-ray emission and the southeast radio relic. Interestingly, there is no X-ray surface brightness peak associated with subcluster two despite a mass $>\text{10}^{\text{15}}$ M$_{\odot}$. This observation provides strong evidence of a merger between subclusters one and two resulting in the ICM of subcluster two becoming substantially disrupted. We will discuss this in more detail in \S \ref{subsec:xdiscuss}. 

There is a 302$\pm$219 $\text{km}\,\text{s}^{-\text{1}}$ LOS velocity difference between these subclusters. The low LOS velocity difference relative to the velocity dispersions follows a pattern for mergers with radio relics; they tend to have LOS velocity differences near and often consistent with zero \citep[see the Merging Cluster Collaboration's radio relic sample paper:][]{sample_paper,Dawson:2014, Ng:2015}. This velocity difference is quite low compared to the expected three dimensional merger velocity based on the masses of the two subclusters. Bimodal mergers in the literature are estimated to have merged with relative velocities of a few thousand km s$^{-\text{1}}$ \citep[see e.g.][]{Dawson:2012, Lage:2015}. The free-fall velocity (or expected maximum relative collision velocity) of subclusters one and two is $\sim$4700 km s$^{-\text{1}}$. Because the LOS velocity difference is so much smaller than the free-fall velocity at pericenter, MACS J1149 is merging near the plane of the sky and/or it is presently viewed near its apocenter. 

\cite{Ogrean:2016} has identified a cold front $\sim$350 kpc to the northeast of the X-ray center that indicates disturbance of the gas in the ICM, and this could be attributed to this merger; however, it is off axis from the supposed merger axis. This would imply the merger between subclusters one and two occurred with a large impact parameter, but this isn't supported by the collinear radio relic and elongation of the ICM. Despite these discrepancies, we can not rule out a connection between this merger and the cold front; however, other physical processes can explain the presence of the cold front as well. Sloshing in the ICM could be one explanation \citep{Roediger:2012}, but these types of cold fronts generally appear in cool-core clusters. Hydrodynamic simulations will be necessary to explore the possibility further, but this is beyond the scope of our analysis. 

Subcluster three, which sits in projection near subcluster one, is an order of magnitude less massive. Both of these subclusters have been matched to a lensing peak identified by \cite{Smith09}. The two lensing peaks are offset by 52\arcsec  (334 kpc at the cluster redshift) and the LOS velocity difference is 3350$\pm$170 km s$^{-\text{1}}$ (see \S\ref{sssec:MCMCGMM}). For comparison, the free-fall velocity is $\sim$ 4370 km s$^{-\text{1}}$. Because the merger LOS velocity difference is so large, the merger is either mostly occurring along the LOS, or the system is unbound and in the background. This gives three distinct possibilities. 

\begin{enumerate}
\item{The LOS velocity difference is attributed to the Hubble flow and the subcluster is not merging. In this case, based on the redshifts, the co-moving distance along the line of sight is 55 Mpc, and the subcluster is not interacting with the rest of MACS J1149.}
\item{The LOS velocity difference is attributed to a peculiar velocity, in which case, subcluster three is merging with subcluster one at a very high velocity along the line of sight. The higher relative LOS velocity for subcluster three indicates that it is in the foreground and not yet at pericenter.}
\item{Same as above, but the merger has already taken place, and subcluster three is just behind subcluster one just after pericenter and shooting into the background along the line of sight.}
\end{enumerate}

\begin{figure}[!htb]
\includegraphics[width=\columnwidth]{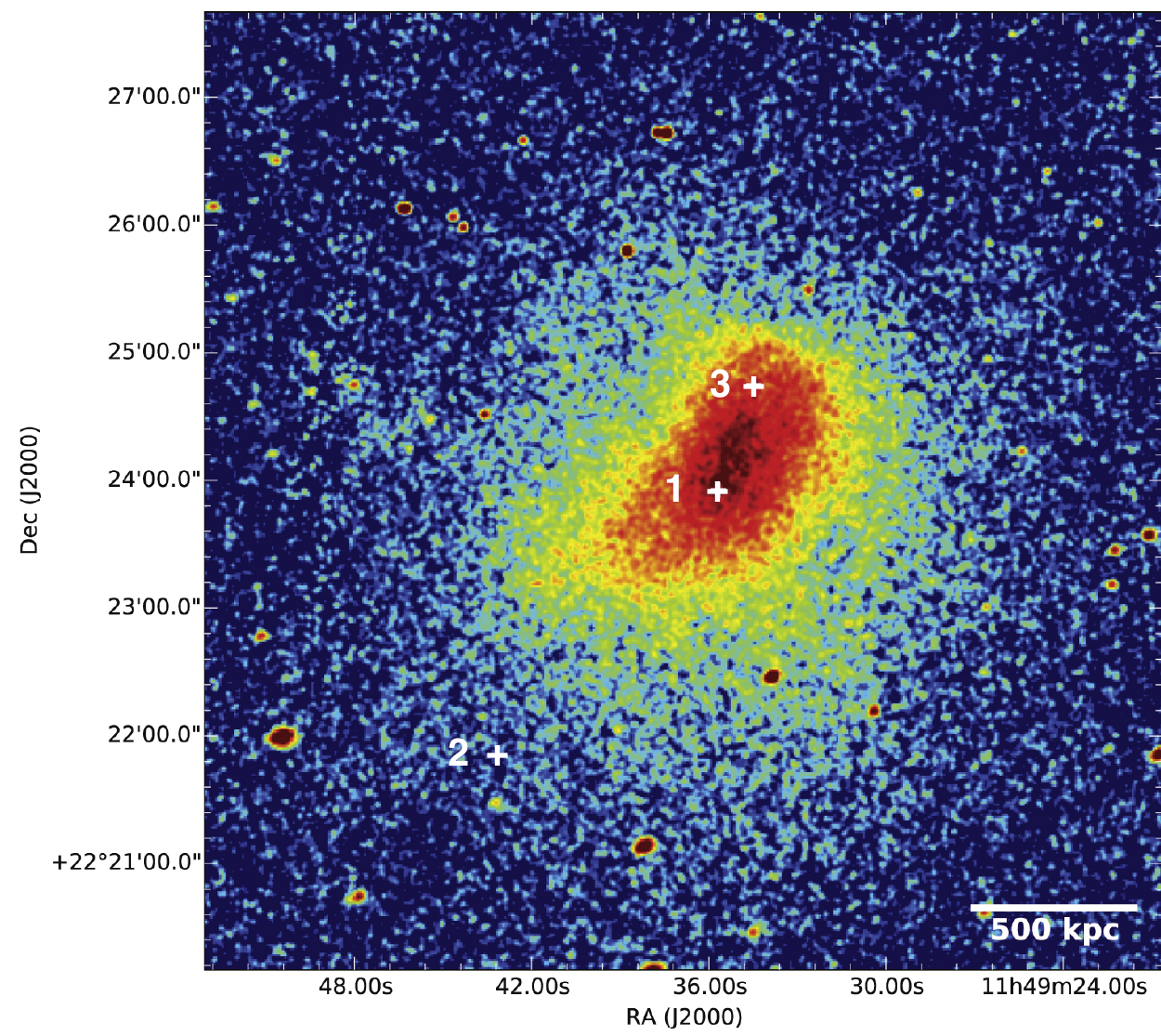}
\caption{Reproduction of Figure 4 of \cite{Ogrean:2016}. The color scale is the \emph{Chandra} surface brightness map in the energy band 0.5-3 keV. The image was exposure and vignetting-corrected and was subtracted of instrumental background before it was smoothed with a Gaussian kernel of width 1$\arcsec$. The three white plusses indicate the locations of subclusters one, two, and three. The X-ray surface brightness peak is extended between to the southeast in the direction of subcluster two as well as toward the northwest in the direction of subcluster three indicating two separate mergers.}
\label{fig:Ogrean_xray}
\end{figure}

To disentangle these scenarios is difficult. Clear evidence of interaction would unambiguously select the third option. Subcluster three has a mass of 1.2$\times$10$^{\text{14}}$ M$_{\odot}$, so we might expect for it to appear in the X-ray surface brightness map as a distinct peak if it were in the foreground prior to merging; however, this may be present and washed out by the ICM of the much more massive subcluster one. On the other hand, we would also expect to see clear merger activity in the ICM if the subcluster had just merged and is past pericenter. Figure \ref{fig:Ogrean_xray} is a reproduction of Figure 4 of \cite{Ogrean:2016} and it shows the location of subcluster one and subcluster three along with the X-ray surface brightness map. The gas is extended in the direction of subcluster three from the main core that is associated with the BCG and subcluster one, but there is no local maximum at the location of subcluster three. Additionally, there is the cold front identified by \cite{Ogrean:2016}, at 96$\%$ confidence, to the northeast of subclusters one and three. The merger between subclusters one and three must be largely along the LOS and very near pericenter, but it is unclear that a LOS merger (perhaps with a southeast--northwest component) could create a cold front to the northeast especially given the small amount of time available since the merger is likely near pericenter.

With the available evidence, we lean toward the scenario that subcluster three has already merged with subcluster one, but we are unable to fully disentangle the possibilities. We note that if subcluster three is merging with subcluster one, it is an order of magnitude less massive, so it would not substantially alter the dynamics of the dominant merger between subclusters one and two. We will explore the dynamics of a possible merger between subclusters one and three in \S\ref{sec:mcmac}. 

\cite{Smith09} identified two additional massive halos in the vicinity of the subcluster one (labeled halos B and E). Each of these are associated with a single bright cluster member galaxy. Halo B is southeast of the BCG and is the second brightest galaxy assigned to the subcluster one. This galaxy is part of the reason that the red sequence luminosity distribution of the northern part of the cluster is elongated along the merger axis and toward the southeast. This could signify that subcluster one is not completely settled. It could also indicate that the merger with subcluster two substantially disturbed subcluster one. Halo E from \cite{Smith09} is associated with a bright galaxy that was assigned to subcluster three. \cite{Smith09} proposed that the presence of multiple peaks is indicative of MACS J1149 being a disassembled cluster. With the added information from the spectroscopic survey, we have shown two of the three additional halos from the \cite{Smith09} strong lensing analysis are associated with a subcluster three, and we have shown the third halo to be part of the same subcluster as the BCG.

In summary, MACS J1149 is comprised of a dominant merger between two extremely massive subclusters. This merger generated the elongated X-ray surface brightness profile and the radio relic in the southeast. There may be an additional merger between subclusters one and three. It has been stated that MACS J1149 is composed of multiple mergers \citep{Smith09, Bonafede12}; however, these studies involve projected information only. Using the spectroscopic data, we have shown it to be composed of a single dominant merger with an additional minor merger possible. With or without the secondary merger, MACS J1149 can be modeled as a bimodal system because subclusters one and two contain 96$\%$ of the mass of the system. We will analyze the dynamics of the merger between subclusters one and two (\S \ref{subsec:primary}) as well as the dynamics of the putative merger between subclusters one and three (\S \ref{subsec:secondary}) in the following section.


\section{Merger Dynamics}\label{sec:mcmac}

\subsection{Merger between subclusters one and two}\label{subsec:primary}

The merging event of MACS J1149 is dominated by the northwest--southeast merger between subclusters one and two from our analysis above. Here we make use of the dynamical analysis code Monte Carlo Merger Analysis Code \citep[MCMAC;][]{Dawson:2012, MCMAC} in order to study this merger. The core assumption of this analysis is two NFW halos merging in otherwise empty space. In this section we aim to estimate the timescale and velocity of the merger.

MCMAC takes five inputs and their associated uncertainties generating Gaussian priors on the following parameters: the projected separation of the two subclusters in the observed state, the masses of the two subclusters, and the redshifts of the two subclusters. In \S\ref{subsec:twod} we found the projected separation between the two peaks of the luminosity weighted galaxy number density map to be 0.99$^{+\text{0.12}}_{-\text{0.15}}$ Mpc. We will use this estimate for the projected separation rather than the distance between the the centers of the blue and green ellipses in Figure \ref{fig:trifit} because we expect the red sequence luminosity to better trace the mass of the system, and the MCMC-GMM analysis treated cluster members equally without accounting for luminosity. The presence of subcluster three overlapping subcluster one may alter the density peak for subcluster one; however, the BCG and other bright central galaxies are members of subcluster one and dominate the luminosity. The luminosity peak is very near the location of the BCG. We refer the reader to Table \ref{table:results} for the mass and redshift inputs for the two subclusters. 

Posterior probability density functions (PDF) for a series of parameters (five input, three geometric, five dynamical) are output. The input parameters consist of: M$_{\text{200-NW}}$, M$_{\text{200-SE}}$, z$_{\text{NW}}$, z$_{\text{SE}}$, and d$_{\text{proj}}$. The geometry parameters consist of the randomly drawn $\alpha$, which is the angle between the plane of the sky and the merger axis ($\alpha=\text{0}$ implies parallel to the plane of the sky), calculated d$_{\text{3D}}$ (the three dimensional separation in the observed state), and calculated d$_{\text{max}}$ (the three dimensional separation at apocenter). The dynamical parameters consist of: TSC$_{\text{0}}$ (time since pericenter in the outbound case), TSC$_{\text{1}}$ (time since pericenter in the returning case), T (time between first and second core passage), v$_{\text{3D}}\left(\text{t}_{\text{col}}\right)$ (three dimensional relative velocity at pericenter) and v$_{\text{3D}}\left(\text{t}_{\text{obs}}\right)$ (three dimensional relative velocity in the observed state).

\begin{figure}[!htb]
\includegraphics[width=\columnwidth]{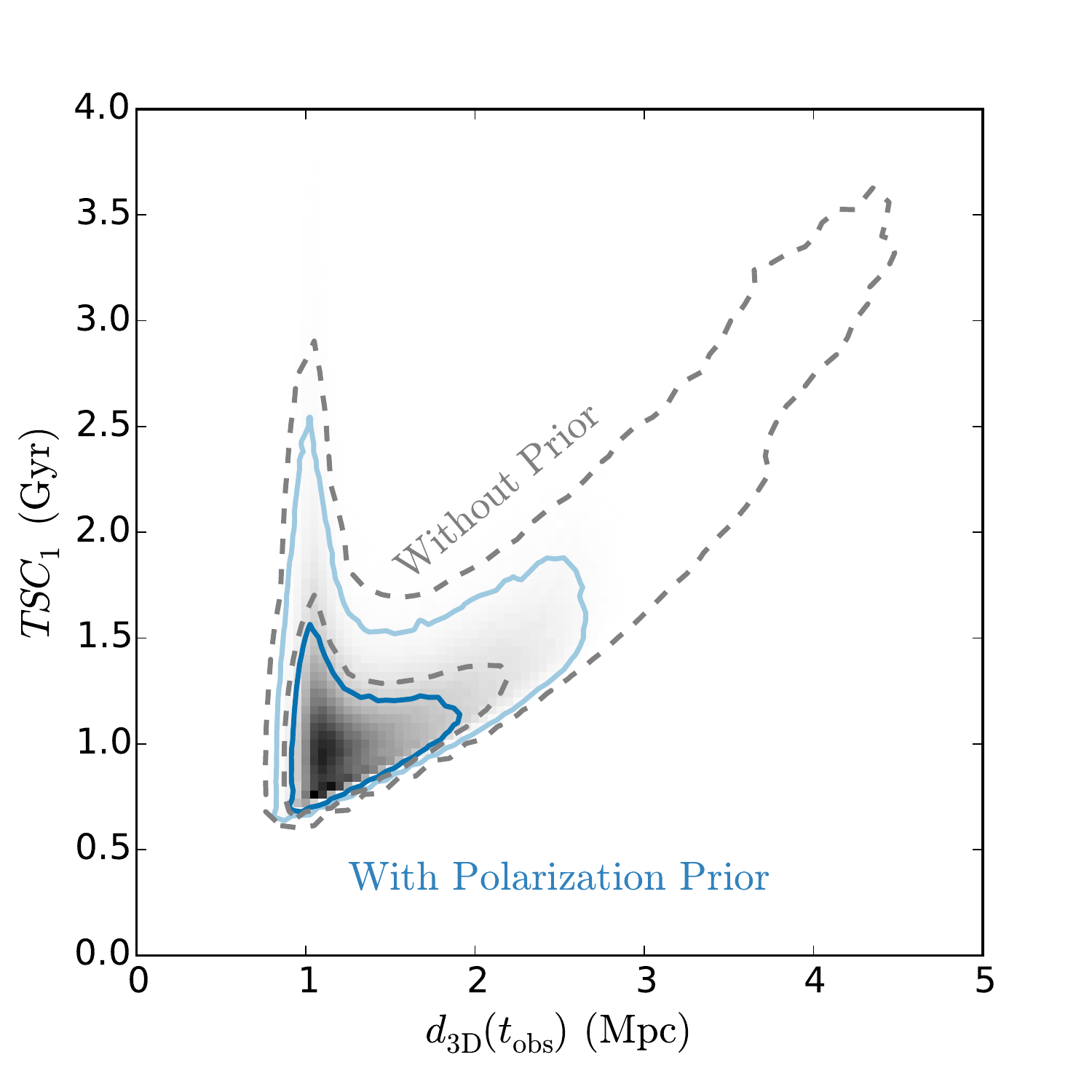}
\caption{MACS J1149 joint posterior PDFs for the time since collision (returning) versus the three dimensional separation in the observed state for the merger between subclusters one and two. The dashed contours show the estimates from the dynamical Monte Carlo analysis without applying any ex post facto priors. The dark and light blue contours show the 68$\%$ and 95$\%$ confidence limits respectively and show the benefit from even the modest prior that the radio relic polarization provides. Observations with JVLA are underway, and the constraint is expected to be stronger.}
\label{fig:alphaprior}
\end{figure}

The largest uncertainty stems from uncertainty in the merger axis angle with the plane of the sky \citep{Dawson:2012}. To constrain $\alpha$, we observe the polarization fraction of the radio emission that corresponds to the radio relics. High radio relic polarization can only be explained by a plane of the sky merger \citep{ensslin1998}. The radio emission from the ICM occurs because of synchrotron emission from relativistic electrons interacting with the magnetic field that is compressed and aligned by a passing shock. Polarized emission suggests a uniform magnetic field, and the polarization fraction depends on the projection of the line of sight onto the magnetic field. A high polarization fraction can only be explained by orthogonal alignment of the LOS and the magnetic field. \cite{Bonafede12} measured a polarization fraction of up to 30$\%$ with a mean polarization fraction of $\sim\,\text{5}\%$ but noted that since beam depolarization likely occurs, this measurement should be treated as a lower limit to the intrinsic polarization. High quality polarization data for the southeast radio relic will offer better constraints when it becomes available. With a better measure of the polarization, uncertainty in the viewing angle could substantially decrease, which will in turn decrease the uncertainty in other dynamical quantities \citep{Dawson:2012,Ng:2015}. We make use of the average polarization fraction of 5$\%$ quoted by \cite{Bonafede12} to constrain $\alpha$. \cite{ensslin1998} presents an analytical model to calculate the maximum observable polarization fraction for a given viewing angle ($\alpha$). Using this model, we find that a 5$\%$ polarization fraction of the radio emission can only be achieved by $\alpha <$ 67$^{\circ}$. This is a modest constraint, but it successfully lowers the uncertainty in many of our estimates. As an example, Figure \ref{fig:alphaprior} shows the effect on the joint posterior PDFs for two parameters in our dynamical analysis. We summarize the output parameters and their 68$\%$ confidence limits (with the polarization prior) in Table ~\ref{table:mcmac}. 

\begin{table}[!htb]
\begin{center}
\caption{MACS J1149 merger parameter estimates including the polarization prior.} 
\begin{tabular}{llll}
Parameter\footnote{\label{parameters}$M_{\text{200}}$ mass; $z$ redshift; $d_{\text{proj}}$ projected subcluster separation; $\alpha$ angle between the merger axis and the plane of the sky; $d_{\text{3D}}$ 4-D subcluster separation; $d_{\text{max}}$ 3-D maximum subcluster separation after core passage; $v_{\text{3D}}\,\left(t_{\text{obs}}\right)$ subcluster relative velocity in the observed state; $v_{\text{3D}}\,\left(t_{\text{col}}\right)$ subcluster relative velocity at core passage; $TSC_{\text{0}}$ time since collision for the outbound scenario;  $TSC_{\text{1}}$ time since collision for the return scenario; $T$ time between collisions. See \cite{Dawson:2012} for more details on these quantities.}	& Location\footnote{\label{biweight}Biweight-statistic location \citep{Beers1990}}	& 68$\%$ LCL--UCL\footnote{\label{bcpcl} Bias-corrected lower and upper confidence limits \citep{Beers1990}}	&   Units\\\hline
M$_{200_{1}}$			& 16.7		& 15.5--18.0			& $10^{14}$\,M$_{\sun}$\\	
M$_{200_{2}}$			& 10.9		& 7.47--14.2			& $10^{14}$\,M$_{\sun}$\\	
$z_{1}$					& 0.54250		& 0.54174--0.54327		& \\	
$z_{2}$					& 0.54106		& 0.54041--0.54171  	& \\	
$d_{\rm proj}$				& 0.995		& 0.931--1.06			& Mpc\\ 	
$\alpha$					& 30			& 11--56				& $^{\circ}$\\ 	
$d_{\rm 3D}$				& 1.15		& 1.00--1.72			& Mpc\\ 	
$d_{\rm max}$				& 1.40		& 1.15--2.09			& Mpc\\ 	
$v_{\rm 3D}(t_{\rm obs})$		& 638		& 245--1710			& km\,s$^{-1}$\\ 	
$v_{\rm 3D}(t_{\rm col})$		& 2770		& 2460--3380			& km\,s$^{-1}$\\ 	
$TSC_{0}$				& 0.683		& 0.426--0.925			& Gyr\\ 	
$TSC_{1}$ 				& 1.16		& 0.913--1.66			& Gyr\\ 	
$T$						& 1.86		& 1.62--2.52			& Gyr\\ 
\end{tabular}
\label{table:mcmac}
\end{center}
\end{table}

Our dynamical model is symmetric under time reversal, so a snapshot of the system is equally described by two scenarios (outbound before apocenter and returning after apocenter). Note that if we accounted for dissipation, each observation still has a degenerate solution, but they are no longer symmetric. To relax the degeneracy, we make use of the radio relic location. The radio relic traces an underlying shock in the ICM from the initial core passage. Simulations of the Bullet Cluster showed the time averaged shock propagation speed to decrease from the merger speed by $\sim$10$\%$ \citep{springel2007}. MACS J1149 is an older merger in either the outbound or return scenarios than the Bullet Cluster, so the simulation does not probe the same time scale. The shock speed in the simulation of the Bullet Cluster is monotonically decreasing. We extrapolated Figure 4 of \cite{springel2007} to the time scale that corresponds to MACS J1149 results in a time averaged shock propagation speed closer to 70$\%$ of the merger speed. We introduce a parameter ($\beta$) to quantify the time averaged decrement of the shock propagation speed. In this manner, the outputs of MCMAC can be used to predict the shock location. These predictions can then be compared to the location of the radio relic to further constrain the time scale. This technique was presented in \cite{Ng:2015}, where it was used to show that El Gordo was in the returning phase.

\begin{figure}[!htb]
\includegraphics[width=\columnwidth]{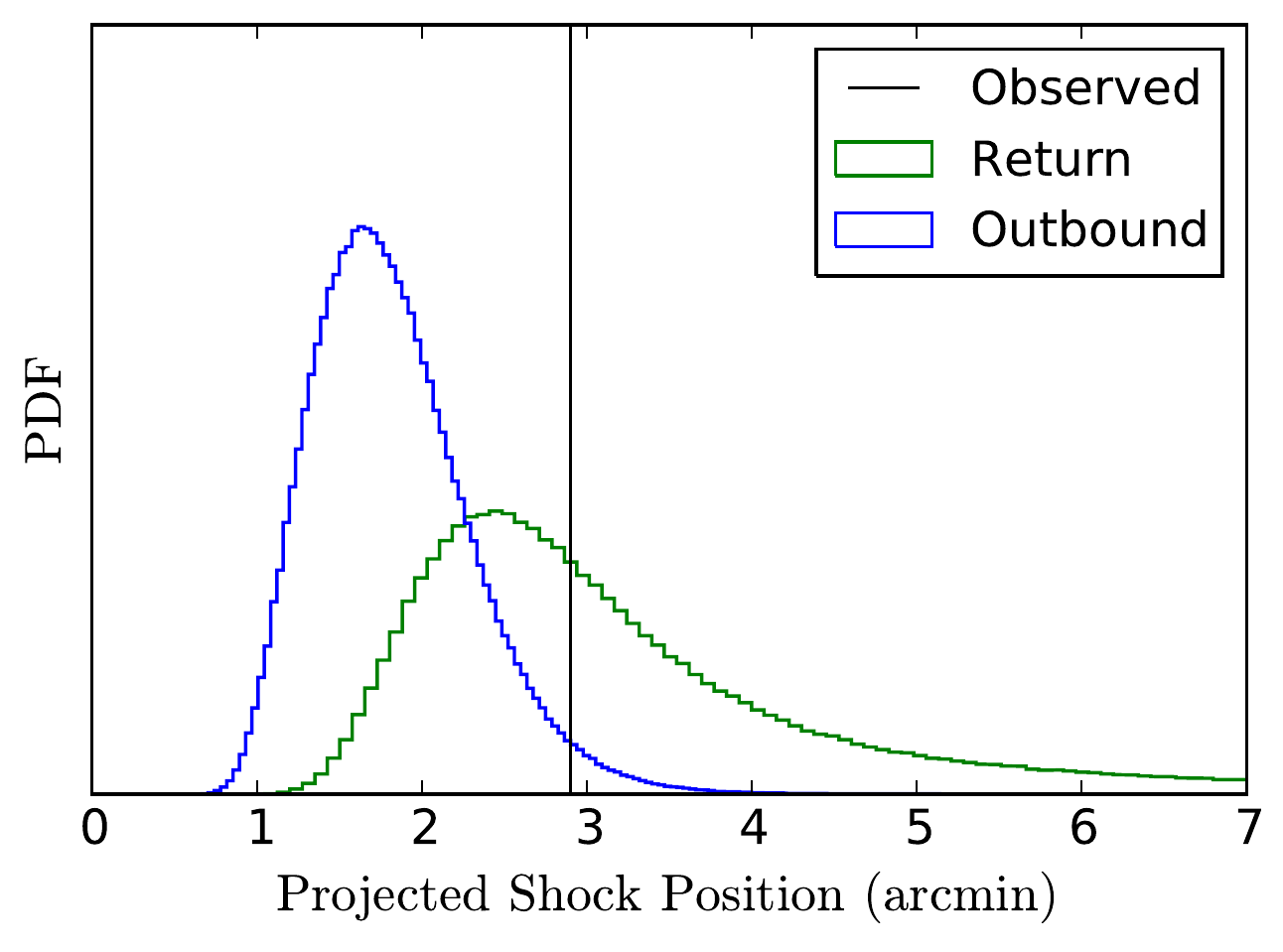}
\caption{Posteriors for the MCMAC predicted position of the southern shock of MACS J1149 in the outbound (blue) and return (green) scenarios in the center of mass frame. The location of the leading edge of the radio relic is shown with the black line. The return scenario is 4.4 times more likely.}
\label{fig:shock}
\end{figure}

For MACS J1149, we assume a time-averaged shock propagation speed of 60-90\% of the 3D collision velocity. Following \cite{Ng:2015}, the position of the shocks in the center of mass frame can be estimated with the MCMAC posteriors:
\begin{equation}\label{eq:outbound}
\text{s}_{\text{i}} = \frac{\text{M}_{\text{j}}}{\text{M}_{\text{i}}+\text{M}_{\text{j}}}\,\beta\,\text{v}_{\text{3D}}\left(\text{t}_{\text{col}}\right)\, \text{TSM}\,\cos{\alpha}
\end{equation}
where $i$ and $j$ refer to the two subclusters, $\beta$ is drawn from $\mathcal{U}\left(\text{0.6},\text{0.9}\right)$, $\text{v}_{\text{3D}}$ is the three dimensional collision velocity of the two subclusters, $\text{TSM}$ is the time since merger pericenter in either the outbound or return case, and $\alpha$ is the angle between the merger axis and the plane of the sky. By this method, the MCMAC posteriors can be used to generate a posterior for the shock locations for the outbound and return scenarios. The estimated shock locations in the outbound and return scenarios are presented in Figure \ref{fig:shock}. We also plot the observed location of the leading edge of the southeast radio relic for comparison. We find the return scenario to be 4.4 times more likely. Thus, we disfavor the outbound time since pericenter estimate (TSM$_{\text{0}}$) in Table \ref{table:mcmac} and accept (with 81$\%$ confidence) TSM$_{\text{1}}$ as the proper estimate of the time since merger pericenter for MACS J1149. In this analysis we assumed a conservative polarization fraction estimate. If we had instead accepted the higher polarization of 30$\%$ from \cite{Bonafede12}, the ratio of likelihood increases to 9.6 (confidence increases to 91$\%$ for the return scenario). This analysis is supported by the location of the X-ray surface brightness peak, which is situated further from the cluster center than the BCG. We will discuss this further in \S \ref{subsec:compare}.

In summary, we have studied the merger between subcluster one and subcluster two and found that they merged with a collision speed of 2770$^{+\text{610}}_{-\text{310}}$ km s$^{-\text{1}}$. In the returning model, core passage occurred 1.16$^{+\text{0.50}}_{-\text{0.25}}$ Gyr before the observed state, which corresponds to a phase of the merger after apocenter and returning for a second collision. We assumed a conservative estimate of the radio relic polarization fraction. A better estimate could further reduce the uncertainty in our dynamical estimates. 

\subsection{Possible merger between subclusters one and three}\label{subsec:secondary}

In \S \ref{sec:scenario}, we explored a few possible interpretations in regards to the placement and dynamical activity of subcluster three with respect to subcluster one. With evidence from the X-ray surface brightness profile (see Figure \ref{fig:Ogrean_xray}), we determined that a scenario where subcluster three has already passed through subcluster one was most likely. 

To explore the chance of such a merger, we implement the analysis presented in \cite{premerger}. The code is a modified version of MCMAC that analyzes all possible scenarios for the subclusters placement including unbound scenarios where subcluster three has a higher redshift because it is far in the background and receding with the Hubble flow. We require the line of sight velocity to be within 500 km s$^{-\text{1}}$ of the Hubble flow velocity for the unbound realizations to be considered valid:
\begin{equation}
V_{\text{LOS}} = V_{\text{LOS,Hubble}} = H\left(\bar{z}\right)\,d_{\text{3d}}\sin\left(\alpha\right)
\end{equation}
where $H\left(\bar{z}\right)$ is the Hubble parameter at the average redshift of the two subclusters, and the other parameters are the same MCMAC parameters defined above. We allow for scatter in the line of sight velocity difference of 500 km s$^{-\text{1}}$ to allow for local peculiar velocities with respect to the Hubble Flow, which show a steep drop off near 500 km s$^{-\text{1}}$ \citep{Bahcall:1996}.

\begin{figure*}[!htb]
\includegraphics[width=\textwidth]{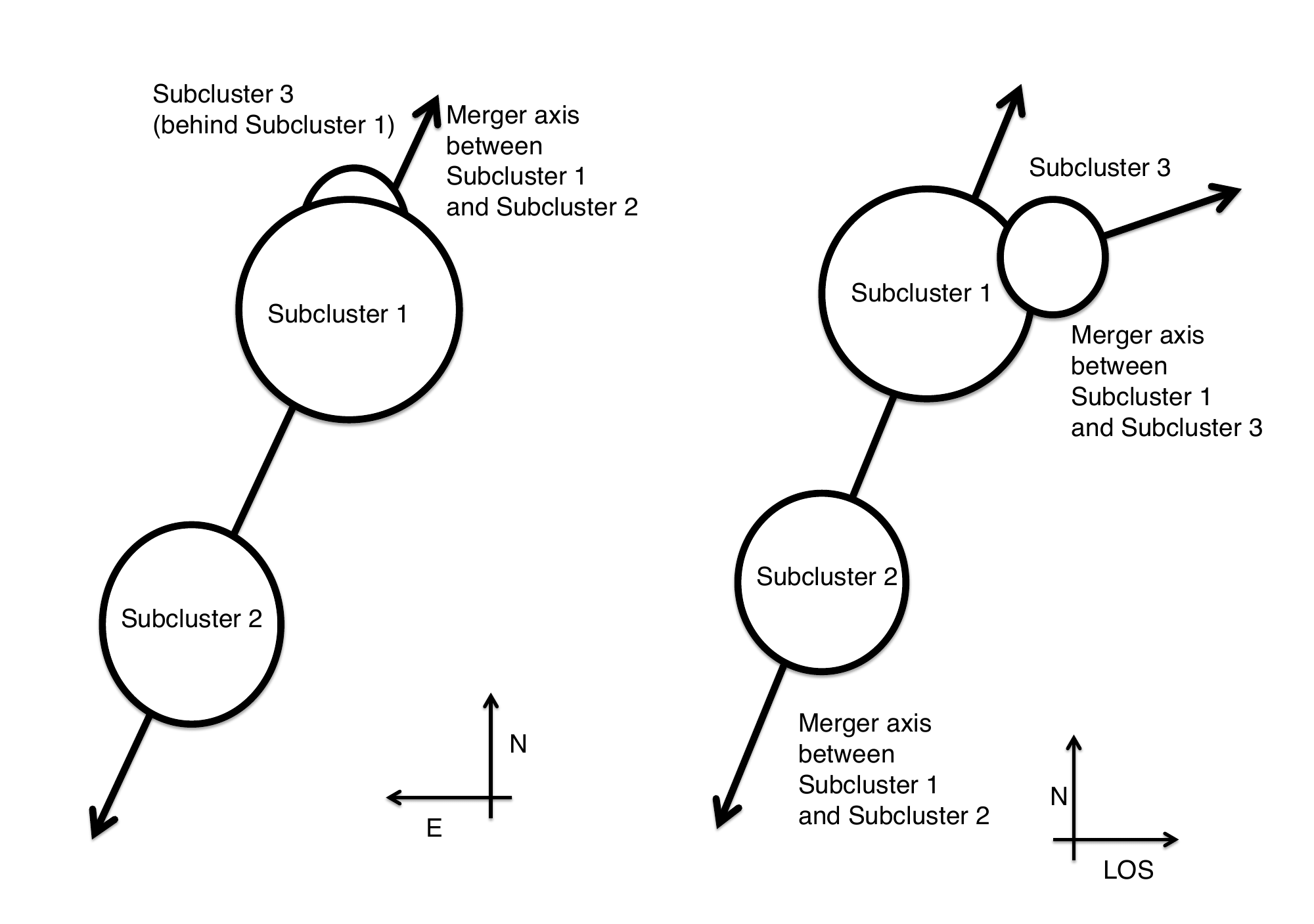}
\caption{Schematic representation MACS J1149. \emph{Left:} as viewed in the observed state with north up and east to the left. The line of sight is into the page. \emph{Right:} as viewed in the observed state but with the viewing angle rotated by 90$^{\circ}$ such that the line of sight is to the right (Earth is to the left), and north is up. The merger axes are approximately to scale with the angles determined from our MCMAC analyses. We assume the bound and post-pericenter scenario for the merger between subclusters one and three.}
\label{fig:scenario}
\end{figure*}

We randomly sample from the observed distributions for the masses, velocities, and projected separation for subclusters one and three and calculate merger parameters (velocities, times, etc) in a similar manner as the standard version of MCMAC. We find that 99.3$\%$ of the realizations correspond to bound scenarios. This indicates that subcluster three is most likely bound to MACS J1149 rather than the velocity difference being associated with the Hubble Flow. For the realizations involving a merger, our dynamical model is agnostic to the current state: we can not distinguish between a state before or after pericenter; however, as discussed in \S\ref{sec:scenario}, there is evidence that subcluster three has already passed through subcluster one (see Figure \ref{fig:Ogrean_xray}). We find that the merger occurred with a three dimensional collision speed of $\sim$4000 km s$^{\text{1}}$ at an angle of $\sim$70$^{\circ}$. The merger occurred recently compared to the merger between subclusters one and two ($\sim$100 Myr before the observed state). In this model, we have assumed the merger was wholly between subclusters one and three, and we have disregarded any effect imparted by the proximity of subcluster two. The merger speed is an underestimate based on our simplified model of the merger; we have ignored the effect of subcluster two on the dynamics of subcluster three. While the putative merger took place between subclusters one and three, subclusters one and two were overlapping at core passage $\sim$1 Gyr before, when subcluster three was accelerating toward collision. We refrain from quantifying uncertainty in our estimates as more detailed simulations are required to understand the three body dynamics. In Figure \ref{fig:scenario}, we present a schematic representation of the subclusters with two viewing angles to help illustrate the preferred scenario from our analysis. 

\section{Discussion}\label{sec:discussion}

\subsection{Results}

MACS J1149 is composed of two dominant subclusters (subclusters one and two) involved in a violent collision that merged with a merger axis near the plane of the sky. A third subcluster (subcluster three) is identified and matched to a lensing peak discussed in the literature. We show subcluster three to be bound and merging with subcluster one in the northern part of MACS J1149. 

The two main subclusters are extremely massive. We estimate their masses from their velocity dispersions to be 16.71$^{+\text{1.25}}_{-\text{1.60}}\times$10$^{\text{14}}$ M$_{\odot}$ and 10.80$^{+\text{3.37}}_{-\text{3.54}}\times$10$^{\text{14}}$ M$_{\odot}$. These two subclusters are assumed to have passed through pericenter because they lie collinearly with the elongation of the X-ray surface brightness distribution along an axis that includes the radio relic to the southeast. Furthermore, the southeast subcluster has been largely stripped of its gas. No X-ray studies have confirmed the presence of a shock associated with the relic, but this is likely due to the lack of sufficient X-ray counts near the location of the relic. We are confident of a link between this relic and the merger of the two main subclusters of MACS J1149 because of the steepening of the radio relic spectral index toward the cluster center and the collinearity with the elongation of the X-ray surface brightness and galaxy distribution. The two main subclusters presently have a LOS relative velocity difference of 302$\pm$219 km s$^{-\text{1}}$ indicating that the merger is occurring with an axis very close to the plane of the sky and/or the merger is very near turnaround (or some combination of the two). Observations of the radio relic polarization fraction constrain the merger axis slightly, and better data should improve the constraint. The current conservative constraint leads to our estimate of the angle between the merger axis and the plane of the sky of 30$^{+\text{25}}_{-\text{19}}$$^{\circ}$. We make use of the location of the radio relic and are able to show that this merger has reached apocenter and is in a returning trajectory with 81$\%$ confidence.

Subcluster three is an order of magnitude less massive than subclusters one and two, and is bound to MACS J1149 and merging with subcluster one with a sharp angle away from the plane of the sky. The merger occurred very recently compared to the merger between subclusters one and two ($\sim$100 Myr ago versus 1160$^{+\text{500}}_{-\text{250}}$ Myr). This secondary merger occurred at an extreme velocity of $\sim$4000 km s$^{-\text{1}}$. Because subcluster three is only 4$\%$ of the total mass of the system, and because the two mergers occurred with a significant amount of time between them, we safely ignore the effect of subcluster three on the merger between subclusters one and two in our analysis of the dynamics of this major dissociative merger.

\subsection{Results of this work in light of results in the literature}\label{subsec:xdiscuss}

Both the X-ray \citep{Ogrean:2016} and radio \citep{Bonafede12} results suggest a merging system with a merger axis projected along a northwest--southeast direction. The radio relic in the southeast shows a steepening trend of the spectral index toward the cluster center, and the elongation of the X-ray surface brightness profile is collinear with this radio relic. Subclusters one and two appear to be the subclusters comprising this merger. The ICM seems to be associated with subcluster one, and the presence of the cluster BCG in this subcluster has led the community to solely study this as the ``core'' of the system. The HST and Chandra observations in the literature were centered around subcluster one, and the added scrutiny of being a Frontier Field cluster has focused the attention on this particular subcluster of MACS J1149. We have shown subcluster two to be similarly massive, which suggests that a similar HST study is warranted to better understand this subcluster. Curiously, there seems to be much less X-ray emission associated with this subcluster and no X-ray surface brightness peak. \cite{Ogrean:2016} recently presented an X-ray study of the cluster with an exposure time of 365 ks, which is more than enough to detect the presence of such a peak if there were one. The lack of an associated X-ray peak provides evidence that this subcluster has passed through subcluster one and was stripped of its gas, which supports the understanding that this merger generated the southeast radio relic. 

We searched the literature for similar scenarios. A massive cluster having no corresponding X-ray surface brightness peak should only be possible in a merging cluster. El Gordo seems to be the most similar cluster, where the northern subcluster lacks an X-ray surface brightness peak. However, there is diffuse emission throughout that appears to be the remains of the associated ICM that has been disrupted by the merger with the cool-core \citep{Menanteau:2012}. Faint, diffuse emission is present in the southeast subcluster of MACS J1149, but it appears to be part of the ICM of the northwest subcluster and it rapidly diminishes toward the southeast; there are no features within several hundred kpc of the center of subcluster two. In another Frontier Field cluster, Abell 2744 \citep{Merten:2011}, the western subcluster seems to lack any X-ray surface brightness peak, but this subcluster is an order of magnitude less massive than the subcluster of interest in MACS J1149 \citep{Merten:2011}. The lack of a X-ray peak in this region of MACS J1149 has led to this subcluster being overlooked since the cluster was first analyzed in detail by \cite{Ebeling:2007}. The elongated weak lensing morphology in Figure 1 of \cite{Umetsu:2014} suggests the presence of this subcluster, but a single mass profile was fit to an elongated shear profile. Several analyses in the literature fit two NFW profiles to bimodal merging clusters \citep[see e.g.][]{Jee:2014, Jee:2015, Jee:2016}. This would allow for more accurate mass estimates for each subcluster. Velocity dispersion mass estimates suggest a massive subcluster, and even though the dynamical mass could bias the mass estimate high, subcluster two would still present quite a massive subcluster without an associated X-ray surface brightness peak.

The X-ray properties presented in \cite{Ogrean:2016} may be translated into total mass estimates for the entire cluster via scaling relations. For the X-ray luminosity, we use Chandra Space Telescope's \emph{PIMMS} \footnote{http://cxc.harvard.edu/toolkit/pimms.jsp} tool to translate the observed flux into a bolometric flux. We then used the \cite{Pratt:2009} scaling relation and estimate a mass of $M_{\text{500}}=\text{12.6}_{-\text{2.2}}^{+\text{1.7}}\times\text{10}^{\text{14}}\,\text{M}_{\odot}$, which translates to $M_{\text{200}}=\text{20.0}_{-\text{2.7}}^{+\text{3.7}}\times\text{10}^{\text{14}}\,\text{M}_{\odot}$ assuming an NFW profile and using the \cite{Duffy} mass--concentration scaling relations. It is still unclear if X-ray luminosity derived masses are over- or underestimates of the true mass in merging clusters. Simulations show a dependance on the viewing angle, and there is significant scatter in actual observations \citep{Takizawa:2010, Zhang:2010}. The X-ray temperature, on the other hand, is a better mass proxy. Using the \cite{Finoguenov:2001} scaling relations and assuming an NFW profile, we find the X-ray temperature scales to a mass of $M_{\text{500}}=\text{15.2}^{+\text{0.62}}_{-\text{0.43}}\times\text{10}^{\text{14}}\,\text{M}_{\odot}$ or $M_{\text{200}}=\text{24.3}^{+\text{10.1}}_{-\text{7.0}}\times\text{10}^{\text{14}}\,\text{M}_{\odot}$, which is in good agreement with the lensing analyses in the literature as well as our dynamical mass estimates. These scaling relations were developed based on a lower redshift sample, and thus only serve as a comparison.

One discrepancy of our merger picture is the presence of the radio relic candidate to the west. Neither merger identified should have created this feature. \cite{Bonafede12} classified this radio feature as a radio relic candidate and noted the presence of a radio and X-ray point source in the vicinity. The potential cold front detected in the existing Chandra data at 96$\%$ confidence \citep{Ogrean:2016} is also unexplained by our two merger scenario. It is unlikely that this cold front is associated with the merger of subclusters one and two. Because it is situated off-axis, a non-zero impact parameter would be required, but the collinearity of the subclusters, the elongation of the ICM, and the radio relic provides evidence against such a scenario. It could also be associated with the merger between subclusters one and three, but that merger is so close to pericenter that there has been very little time for the cold front to propagate to its observed position. The presence of cold fronts in the ICM is not unique to merging systems. For example, \cite{Roediger:2012} showed in simulations that sloshing in the ICM of mergers was capable of creating cold fronts. Furthermore, \cite{AandM:2006} observed that even relaxed clusters have cold fronts due to sloshing in the ICM. However, we cannot rule out the possibility of a merger induced cold front in this case.

\subsection{Velocity dispersion mass estimates}\label{subsec:vdisp_mass}

The use of velocity dispersion derived mass estimates will undoubtedly raise concerns. It is a well known fact that these masses are biased high in mergers due to the departure of the system from its virialized status \citep{Takizawa:2010, Saro:2013}. \cite{Ogrean:2016} suggest that the X-ray surface brightness profile is relatively regular compared to other major mergers, which could imply an older merger. This is substantiated by our dynamical analysis that showed the merger occurred 1.16$^{+\text{0.50}}_{-\text{0.25}}$ Gyr before the observed state. \cite{Pinkney:1996} presents a 3:1 mass ratio merger, where the velocity dispersion bias substantially diminishes by 1 Gyr after core passage. In the outbound scenario, MACS J1149 may have a substantially biased velocity dispersion; however, in the return scenario, which is preferred based on the radio relic location, the bias is small suggesting that the velocity dispersion masses are accurate. A second form of bias is introduced by the misidentification of galaxies to the wrong subclusters, which artificially inflates the velocity dispersion and thus the mass estimate. Our MCMC-GMM code is designed to separate contaminants from overlapping Gaussians, and the large LOS velocity difference between subcluster one and subcluster three supports a clean separation. In another bimodal system with a similar physical configuration and analysis scheme, \cite{Dawson:2014} found dynamical mass estimates that are biased high by $\sim$40$\%$ from the weak lensing mass estimates of \cite{Jee:2014}. Even with these biases, \cite{Dawson:2012} showed that the results for the Monte Carlo analysis used in this paper to study the merger dynamics are relatively insensitive to the mass input. To test this, we draw from from the Monte Carlo chains and apply an additional Gaussian weight to form a distribution of the outputs that has a 40$\%$ reduction in the masses. The dynamical parameters (velocity and time) each changed by less than 10$\%$ despite the 40$\%$ change in mass. Each output (except the masses) stayed within the 68$\%$ confidence limits in Table \ref{table:mcmac}. 

Still, a better estimate of the masses of the two subclusters could be achieved from a joint strong and weak lensing analysis. The CLASH collaboration already analyzed MACS J1149 in this manner \citep{Umetsu:2015}; however, they fit a single NFW profile centered on the cD galaxy at the center of subcluster one. This resulted in an estimate of M$_{\text{200}}$ of 25.02 $\pm$ 5.53 $\times$ 10$^{\text{14}}$ M$_{\odot}$ with a concentration of 2.1 $\pm$ 0.6 and a scale radius of 1.12 $\pm$ 0.35 Mpc. This results in an estimate for $R_{\text{200}}$ of 2.35 Mpc. The projected separation of the northwest and southeast subclusters is 0.99$^{+\text{0.12}}_{-\text{0.15}}$ Mpc. The background subcluster also lies within this value of $R_{\text{200}}$. The combined masses from velocity dispersions for these three subclusters is 28.71$^{+\text{4.81}}_{-\text{5.48}}$ $\times$ 10$^{\text{14}}$ M$_{\odot}$, which is in good agreement with the lensing study. Figure 1 of \cite{Umetsu:2014} presents a low resolution mass map of MACS J1149 that clearly shows subcluster one with a clear extension in the direction of subcluster two. The core of subcluster one has been studied with strong lensing, which resulted in an estimate for the mass within 500 kpc of 6.7 $\pm$ 0.4 $\times$ 10$^{\text{14}}$ M$_{\odot}$ indicating a very dense core \citep{Smith09}. No analogous study has been completed for the southeast subcluster because it lacks HST data. Our results show a very massive subcluster and provide motivation for further observations in this region.

\subsection{Comparison to other MCMAC analyses}\label{subsec:compare}

The merger between subclusters one and two of MACS J1149 is the oldest cluster merger yet modeled with MCMAC. Other systems studied include the Bullet Cluster \citep{Dawson:2012} and El Gordo \citep{Ng:2015}. The Bullet Cluster is a much younger merger and is in a much earlier phase of the merger \citep{Dawson:2012}. El Gordo, on the other hand, was also shown to be returning for a second core-passage \citep{Ng:2015}. Each of these three systems involve extremely massive clusters and violent mergers, but only MACS J1149 is composed of more than two subclusters.
\begin{figure}[!htb]
\includegraphics[width=\columnwidth]{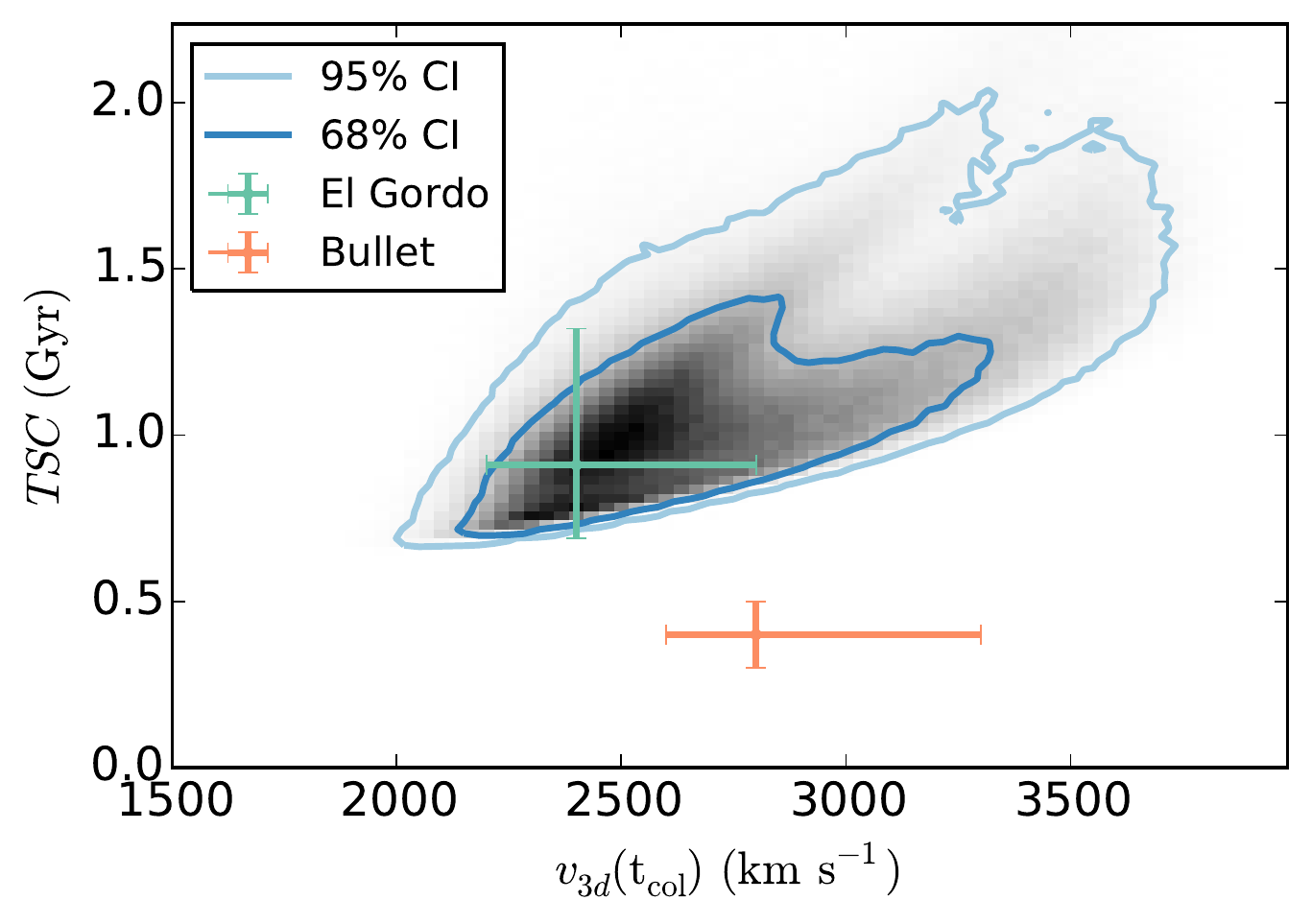}
\caption{The joint posterior PDFs for time since collision and three dimensional merger velocity of the main merger between subclusters one and two of MACS J1149. Dark blue and light blue contours represent 68\% and 95\% confidence respectively. The values for the same parameters for the Bullet Cluster \citep{Dawson:2012} and El Gordo \citep{Ng:2015} are overlaid for comparison.}
\label{fig:TSM1compare}
\end{figure}
Figure \ref{fig:TSM1compare} shows that the merger between subclusters one and two of MACS J1149 is dynamically similar to El Gordo. Both mergers are extremely massive and are in the returning phase of their merger, and have two extremely massive subclusters. The similarity with El Gordo is solidified by comparing the X-ray surface brightness profile of the two clusters. MACS J1149 has an X-ray peak slightly leading subcluster one, which is similar to the eastern subcluster of El Gordo \citep[see Figure \ref{fig:rgb} above and Figure 1 of][]{Ng:2015} where the lensing derived mass peak is lagging the X-ray cool-core remnant \citep{Menanteau:2012, JeeGordo}. \cite{Smith09} shows the X-ray peak in MACS J1149 to be offset from the lensing mass peak by 15$\arcsec$ along the merger axis to the northwest (i.e. further from the center of mass of the cluster). During a merger, we expect the X-ray peak to lag the mass peak for a time; however, it then is reaccelerated forward by the slingshot effect. This effect was first seen in simulations by \cite{AandM:2006}, described by \cite{MandV:2007}, and seen observationally in another Frontier Field cluster: Abell 2744 \citep{Owers:2011, Merten:2011}. This scenario could describe the observations of MACS J1149 and supports the returning scenario, which is consistent with the relatively regular state in the X-ray surface brightness \citep{Ogrean:2016} and the position of the southeast radio relic. However, the secondary merger between subclusters one and three could have also caused this offset. This merger is more similar to the the Bullet Cluster, which has a much less massive subcluster merging with an extremely massive one. The Bullet Cluster merger is occurring in the plane of the sky, and the analogous merger in MACS J1149 is occurring largely along the line of sight.

This is the first analysis containing two separate mergers studied with MCMAC. Generally speaking, this type of system should be studied with more intensive N-body and hydrodynamical simulations. Our dynamical modeling provides a good set of initial conditions for such studies.

\subsection{Summary}

\begin{itemize}
\item{MACS J1149 is an extremely massive cluster composed of three subclusters. Subcluster one is in the northwest of the cluster and has a mass of $M_{\text{200}}=\text{16.7}^{+\text{1.25}}_{-\text{1.60}}\times\text{10}^{\text{14}}$ M$_{\odot}$; subcluster two is in the southeast and has a mass of $M_{\text{200}}=\text{10.8}^{+\text{3.37}}_{-\text{3.54}}\times\text{10}^{\text{14}}$ M$_{\odot}$; subcluster three is very close to subcluster one and has a mass of $M_{\text{200}}=\text{1.20}^{+\text{0.19}}_{-\text{0.34}}\times\text{10}^{\text{14}}$ M$_{\odot}$.}
\item{Subcluster two has been previously unidentified. This is likely due to the lack of an associated X-ray surface brightness peak. It lacks the HST/ACS observations that enabled the strong lensing analysis in the northern part of the cluster, but there is additional evidence of its existence in weak lensing studies using Subaru/SuprimeCam in the literature.}
\item{The two more massive subclusters had a first pericenter passage 1.16$_{-\text{0.25}}^{+\text{0.50}}$ Gyr ago with a collision speed of 2770$^{+\text{610}}_{-\text{310}}$ km s$^{-\text{1}}$ and have reached apocenter and turned around for a second core passage. This merger created a radio relic and stripped the southeast subcluster of its gas.}
\item{The primary merger is dynamically similar to El Gordo in terms of its merger speed and time since pericenter.}
\item{The northern subcluster is presently involved in a merger with the smallest subcluster, which impacted at a sharp angle with respect to the plane of the sky merged with a extreme velocity approximately 1 Gyr after the primary merger took place.}
\item{The ICM is primarily associated with subcluster one, but it is peaked ahead of the BCG. Explanations for this observation include a slingshot effect scenario, where the gas has catapulted beyond the BCG as the subclusters reached apocenter, or the recent merger between subclusters one and three could have disturbed the ICM in this direction. It is difficult to select between these scenarios, and it is possible that it is actually a combination of the two.}
\end{itemize}

\section{Acknowledgments}
We would like to thank the broader membership of the Merging Cluster Collaboration for their continual development of the science motivating this work. We would like to specifically thank Bryant Benson for his diligent debugging of the MCMC-GMM code.\\
This material is based upon work supported by the National Science Foundation under Grant No. (1518246).
\\
Part of this was work performed under the auspices of the U.S. DOE by LLNL under Contract DE-AC52-07NA27344.
\\
GO acknowledges support by NASA through a Hubble Fellowship grant HST-HF2-51345.001-A awarded by the Space Telescope Science Institute, which is operated by the Association of Universities for Research in Astronomy, incorporated under NASA contract NAS5-26555. 
\\
RJW is supported by a Clay Fellowship awarded by the Harvard-Smithsonian Center for Astrophysics.
\\
The authors would like to give special thanks to the CLASH and the MACS surveys for making their data public.
\\
This research has made use of the NASA/IPAC Extragalactic Database (NED) which is operated by the Jet Propulsion Laboratory, California Institute of Technology, under contract with the National Aeronautics and Space Administration.
\\
This research has made use of NASA's Astrophysics Data System.
\\
This research made use of Montage. It is funded by the National Science Foundation under Grant Number ACI-1440620, and was previously funded by the National Aeronautics and Space Administration's Earth Science Technology Office, Computation Technologies Project, under Cooperative Agreement Number NCC5-626 between NASA and the California Institute of Technology.
\\
This research made use of APLpy, an open-source plotting package for Python hosted at http://aplpy.github.com.
\\
{\it Facilities:} \facility{Keck II (DEIMOS)}, \facility{Subaru (Suprime-Cam)}, \facility{GEMINI-N (GMOS)}, \facility{Chandra (ACIS-I)}, \facility{WSRT}, \facility{GMRT}, \facility{VLA}.

\bibliographystyle{mn2e}
\bibliography{mcc}

\end{document}